\documentclass[prl,twocolumn,superscriptaddress]{revtex4-1}
\usepackage{}
\usepackage{amssymb}
\usepackage{amsfonts}
\usepackage{bbm}
\usepackage{mathrsfs}
\usepackage{graphics,graphicx,epsfig,bm,amsmath,amsthm,amssymb}
\usepackage{bm}
\usepackage{bbm}
\usepackage{longtable}
\usepackage{multirow}
\usepackage{array}
\usepackage{color}
\usepackage[usenames,dvipsnames]{xcolor}

\usepackage{float}
\usepackage[a4paper,colorlinks=true,
linkcolor=blue,citecolor=blue,
pdfauthor={ },
pdftitle={ },
pdfsubject={ },
pdfkeywords={ }]{hyperref}

\begin{document}

\title{Observation of parity-time symmetry breaking in a single spin system}

\author{Yang Wu}
\thanks{These authors contributed equally to this work.}
\affiliation{Hefei National Laboratory for Physical Sciences at the Microscale, University of Science and Technology of China, Hefei 230026, China}
\affiliation{CAS Key Laboratory of Microscale Magnetic Resonance and Department of Modern Physics, University of Science and Technology of China, Hefei 230026, China}
\affiliation{Synergetic Innovation Center of Quantum Information and Quantum Physics, University of Science and Technology of China, Hefei 230026, China}

\author{Wenquan Liu}
\thanks{These authors contributed equally to this work.}
\affiliation{Hefei National Laboratory for Physical Sciences at the Microscale, University of Science and Technology of China, Hefei 230026, China}
\affiliation{CAS Key Laboratory of Microscale Magnetic Resonance and Department of Modern Physics, University of Science and Technology of China, Hefei 230026, China}
\affiliation{Synergetic Innovation Center of Quantum Information and Quantum Physics, University of Science and Technology of China, Hefei 230026, China}

\author{Jianpei Geng}
\affiliation{3rd Physikalisches Institut, University of Stuttgart, Pfaffenwaldring 57, 70569 Stuttgart, Germany}

\author{Xingrui Song}
\affiliation{Hefei National Laboratory for Physical Sciences at the Microscale, University of Science and Technology of China, Hefei 230026, China}

\author{Xiangyu Ye}
\affiliation{Hefei National Laboratory for Physical Sciences at the Microscale, University of Science and Technology of China, Hefei 230026, China}
\affiliation{CAS Key Laboratory of Microscale Magnetic Resonance and Department of Modern Physics, University of Science and Technology of China, Hefei 230026, China}
\affiliation{Synergetic Innovation Center of Quantum Information and Quantum Physics, University of Science and Technology of China, Hefei 230026, China}

\author{Chang-Kui Duan}
\affiliation{Hefei National Laboratory for Physical Sciences at the Microscale, University of Science and Technology of China, Hefei 230026, China}
\affiliation{CAS Key Laboratory of Microscale Magnetic Resonance and Department of Modern Physics, University of Science and Technology of China, Hefei 230026, China}
\affiliation{Synergetic Innovation Center of Quantum Information and Quantum Physics, University of Science and Technology of China, Hefei 230026, China}

\author{Xing Rong}
\email{xrong@ustc.edu.cn}
\affiliation{Hefei National Laboratory for Physical Sciences at the Microscale, University of Science and Technology of China, Hefei 230026, China}
\affiliation{CAS Key Laboratory of Microscale Magnetic Resonance and Department of Modern Physics, University of Science and Technology of China, Hefei 230026, China}
\affiliation{Synergetic Innovation Center of Quantum Information and Quantum Physics, University of Science and Technology of China, Hefei 230026, China}

\author{Jiangfeng Du}
\email{djf@ustc.edu.cn}
\affiliation{Hefei National Laboratory for Physical Sciences at the Microscale, University of Science and Technology of China, Hefei 230026, China}
\affiliation{CAS Key Laboratory of Microscale Magnetic Resonance and Department of Modern Physics, University of Science and Technology of China, Hefei 230026, China}
\affiliation{Synergetic Innovation Center of Quantum Information and Quantum Physics, University of Science and Technology of China, Hefei 230026, China}

\begin{abstract}
A fundamental axiom of quantum mechanics requires the Hamiltonians to be Hermitian which guarantees real eigen-energies and probability conservation.
However, a class of non-Hermitian Hamiltonians with Parity-Time ($\mathcal{PT}$) symmetry can still display entirely real spectra \cite{PRL_Bender_1998}.
The Hermiticity requirement may be replaced by $\mathcal{PT}$ symmetry to develop an alternative formulation of quantum mechanics\cite{PRL_Bender_2002, RPP_Bender}.
A series of experiments have been carried out with \emph{classical} systems including optics\cite{NP_Ruter}, electronics\cite{PRL_N_Bender, Nature_Assawaworrarit, NC_Choi}, microwaves\cite{PRL_Bittner}, mechanics\cite{AJP_Bender} and acoustics\cite{PRX_Zhu, NC_Popa, NC_Fleury}.
However, there are few experiments to investigate $\mathcal{PT}$ symmetric physics in quantum systems.
Here we report the first observation of the $\mathcal{PT}$ symmetry breaking in a single spin system.
We have developed a novel method to dilate a \emph{general} $\mathcal{PT}$ symmetric Hamiltonian into a Hermitian one, which can be realized in a practical quantum system.
Then the state evolutions under $\mathcal{PT}$ symmetric Hamiltonians, which range from $\mathcal{PT}$ symmetric unbroken to broken regions, have been experimentally observed with a single nitrogen-vacancy (NV) center in diamond.
Due to the universality of the dilation method, our result opens a door for further exploiting and understanding the physical properties of $\mathcal{PT}$ symmetric Hamiltonian in quantum systems.

\end{abstract}

\maketitle
In quantum mechanics, the real energies of a system are guaranteed by a fundamental axiom associated with the Hermiticity of physical observables.
However, a class of non-Hermitian Hamiltonians satisfying $\mathcal{PT}$ symmetry can still exhibit real eigenenergies\cite{PRL_Bender_1998}.
In principle, wider range of systems can be described by $\mathcal{PT}$ symmetric Hamiltonians comparing to Hermitian ones.
A Hamiltonian $H$ is considered to be $\mathcal{PT}$ symmetric if $[\mathcal{PT}, H] = 0$, where $\mathcal{P}$ and $\mathcal{T}$ denote the parity and time-reversal operators, respectively.
A sufficient condition for a $\mathcal{PT}$ symmetric Hamiltonian $H$ to exhibit entire real eigenvalues is that $H$ corresponds to the region of unbroken $\mathcal{PT}$ symmetry, where any eigenfunction of $H$ is simultaneously an eigenfunction of the $\mathcal{PT}$ operator.
Otherwise, if the $\mathcal{PT}$ symmetric Hamiltonian $H$ and the $\mathcal{PT}$ operator possess different eigenfunctions, $H$ corresponds to the region of broken $\mathcal{PT}$ symmetry.
A Hamiltonian $H$ with a broken $\mathcal{PT}$ symmetry is typically associated with the presence of complex eigenenergies.
An alternative formulation of quantum mechanics can be established in which the axiom of Hermiticity is replaced by the condition of $\mathcal{PT}$ symmetry\cite{PRL_Bender_2002, RPP_Bender}.

The rich physics associated with $\mathcal{PT}$ symmetric Hamiltonian have aroused considerable experimental interest\cite{NP_El_Ganainy}.
A series of experiments have been performed with \emph{classical} approaches.
The optical analog of $\mathcal{PT}$ symmetric quantum mechanics was firstly proposed\cite{NP_Ruter}, then the concept was quickly extended to other systems, such as electronics\cite{PRL_N_Bender, Nature_Assawaworrarit, NC_Choi}, microwaves\cite{PRL_Bittner}, mechanics\cite{AJP_Bender}, acoustics\cite{PRX_Zhu, NC_Popa, NC_Fleury}, and optical systems with atomic media\cite{PRL_Hang, NP_Peng_2016, PRL_Zhang}.
In particular, the experimental research on $\mathcal{PT}$ symmetric classical optical systems has been honored as a most important accomplishment in the past decade\cite{NP_Cham} and has stimulated many applications such as unidirectional light transport\cite{Science_Feng_2011, NP_Peng_2014} and single-mode lasers\cite{Science_Feng_2014, Science_Hodaei}.
As a contrast, it is still of challenge to experimentally investigate
$\mathcal{PT}$ symmetric Hamiltonian related physics in quantum systems.
This is because that experimental quantum systems are governed by Hermitian Hamiltonians.
A possible approach is to realize a $\mathcal{PT}$ symmetric Hamiltonian in an open quantum system, but it is generally difficult to realize a controllable $\mathcal{PT}$ symmetric Hamiltonian by controlling the environment\cite{PRA_Gardas}.
Some progress has been made with this approach in the system of light-matter quasiparticles\cite{Nature_Gao, NC_Zhang}.
A lossy Hamiltonian has been constructed to simulate the quantum
dynamics under a corresponding $\mathcal{PT}$ symmetric Hamiltonian\cite{arxiv_Li}.
In other experiments, non-unitary operators are designed to circumvent engineering $\mathcal{PT}$ symmetric Hamiltonians\cite{NPhoton_Tang, NP_Xiao}.
Recently, an approach has been developed to dilate a single-qubit $\mathcal{PT}$ symmetric Hamiltonian into a Hermitian Hamiltonian in a higher dimensional Hilbert space\cite{PRL_Gunther}.
This method was further developed for dilation of arbitrary-dimensional Hamiltonians\cite{PRL_Kawabata}.
However, these methods can not be utilized to dilate the $\mathcal{PT}$ broken Hamiltonian. Thus to observe the broken of $\mathcal{PT}$ symmetry in a single quantum system, such as a single spin, remains elusive.

In this paper, we report the first observation of the broken of $\mathcal{PT}$ symmetry in a single spin system.
We develop a universal method to dilate a \emph{general} $\mathcal{PT}$-symmetric Hamiltonian into a Hermitian one with an ancilla.
This method is capable of Hermitian dilation of  general $\mathcal{PT}$ symmetric Hamiltonian with arbitrary dimension, while only one ancilla qubit is required.
A single nitrogen-vacancy center in diamond has been utilized as a platform to demonstrate our method.
Both the state evolutions under  $\mathcal{PT}$ symmetric broken and unbroken Hamiltonians have been successfully observed.

We consider a quantum system, $s$, which is driven by a $\mathcal{PT}$ symmetric Hamiltonian $H_s$.
The quantum state of $s$ is denoted by $|\psi(t)\rangle$, which satisfies the Schr$\ddot{\text{o}}$dinger type equation, $i\frac{d}{dt}|\psi(t)\rangle=H_s|\psi(t)\rangle$.
To realize $H_s$ in a quantum system, an ancilla qubit $a$ is introduced to dilate $H_s$ into a Hermitian Hamiltonian $H_{s,a}(t)$.
The state of the combined system, $|\Psi(t)\rangle$, is a dilation of $|\psi(t)\rangle$ with the form
\begin{equation}
|\Psi(t)\rangle=|\psi(t)\rangle|-\rangle+\eta(t)|\psi(t)\rangle|+\rangle,
\end{equation}
where $|-\rangle=(|0\rangle-i|1\rangle)/\sqrt{2}$ and $|+\rangle=-i(|0\rangle+i|1\rangle)/\sqrt{2}$ are the eigenstates of $\sigma_y$ forming an orthonormal basis of the ancilla qubit and $\eta(t)$ is a linear operator.
When a measurement is applied on the ancilla qubit and $|-\rangle$ is postselected, the evolution of quantum state, $|\psi(t)\rangle$, driven by  $\mathcal{PT}$ symmetric Hamiltonian $H_s$ is produced.

Now the key is to derive the expression of $H_{s,a}(t)$.
The evolution governed by the Hermitian Hamiltonian $H_{s,a}(t)$ can be described by the Schr$\ddot{\text{o}}$dinger equation,
\begin{equation}
i\frac{d}{dt}|\Psi(t)\rangle=H_{s,a}(t)|\Psi(t)\rangle.
\end{equation}
The Hamiltonian, $H_{s,a}(t)$, can be designed flexibly according to practical physical systems for the reason that $H_{s,a}(t)$ is not uniquely determined (see Supplementary Material for details).
For example, $H_{s,a}(t)$ can be designed to be
\begin{equation}
H_{s,a}(t) = \Lambda(t) \otimes I + \Gamma(t) \otimes \sigma_z,
\end{equation}
where
$\Lambda(t)=\{H_s(t)+[i\frac{d}{dt}\eta(t)+\eta(t)H_s(t)]\eta(t)\}M^{-1}(t)$, and $\Gamma(t)=i[H_s(t)\eta(t)-\eta(t)H_s(t)-i\frac{d}{dt}\eta(t)]M^{-1}(t)$.
The time-dependent operator $M(t)$ have the form $M(t)=\eta^{\dagger}(t)\eta(t)+I$, where
$\sigma_x$, $\sigma_y$ and $\sigma_z$ are Pauli operators and $I$ is the identity matrix.
This derivation of $H_{s,a}(t)$ holds for \emph{arbitrary} Hamiltonians $H_s$
(see Supplementary Material for the proof).
Our method can be utilized to Hermitianly dilate a \textit{general} $\mathcal{PT}$ symmetric Hamiltonian. Thus it paves a way to a direct experimental investigate \textit{general} $\mathcal{PT}$ symmetric related physics in quantum systems.

For clarity and without loss of generality, the $\mathcal{PT}$ symmetric Hamiltonian with the form,
\begin{equation}
H_s = \left[ \begin{array}{cc}
ir & 1\\
1  & -ir
\end{array}
\right ],
\end{equation}
is taken as an example, where $r$ is a real number.
The eigenvalues of $H_s$ are $E=\pm\sqrt{1-r^2}$.
In the region $|r|<1$, the eigenvalues $E$ are real and the system is in a unbroken-symmetry region.
Especially, when $r=0$, the Hamiltonian $H_s$ is Hermitian.
When $|r|>1$, the imaginary part of $E$ appears and the system is in a broken-symmetry region.
The point $|r|=1$ is known as the exceptional point.
The dilated Hermitian Hamiltonian of the $H_s$ can be derived from equation (3) by taking $\eta(t) = \eta_0(t) I$, where $\eta_0(t)$ is a real parameter.
By expanding $\Lambda(t)$ and $\Gamma(t)$ in terms of Pauli operators, the dilated Hamiltonian has the form (see Supplementary Material for details)
\begin{equation}
\begin{aligned}
H_{s,a}(t) &= A_1(t)\sigma_x \otimes I + A_2(t)I \otimes \sigma_z\\
&+A_3(t)\sigma_y \otimes \sigma_z +  A_4(t)\sigma_z \otimes \sigma_z,
\end{aligned}
\end{equation}
where $A_1(t)$, $A_2(t)$, $A_3(t)$ and $A_4(t)$ are real parameters corresponding to the $\mathcal{PT}$ symmetric Hamiltonian $H_s$ shown in equation (4).

\begin{figure}
\centering
\includegraphics[width=0.9\columnwidth]{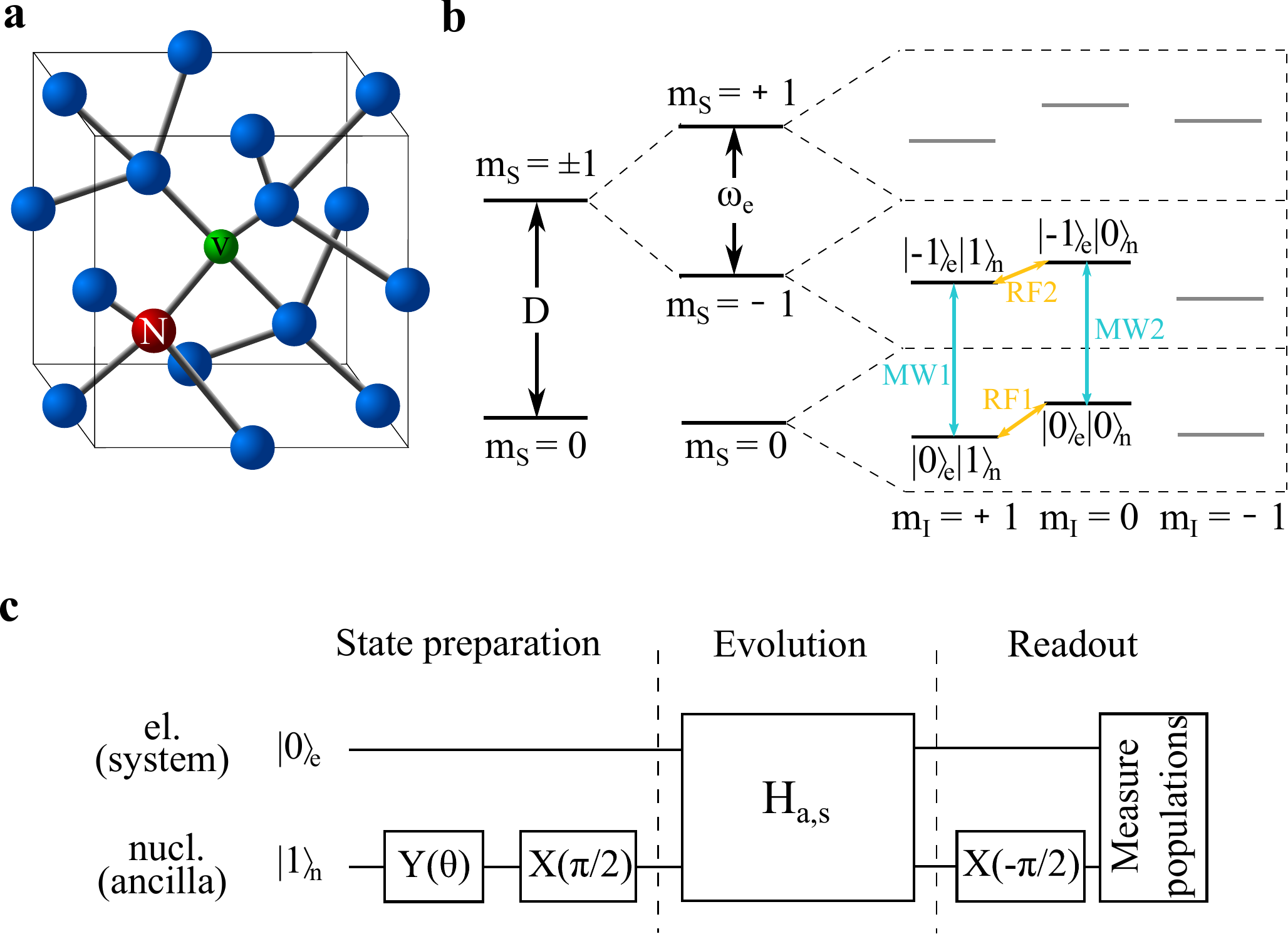}
    \caption{Constructing of $\mathcal{PT}$ symmetric Hamiltonian in NV center. \textbf{a}, Schematic atomic structure and energy levels of the NV center. \textbf{b}, Hyperfine structure of the coupling system with NV electron spin and $^{14}$N nuclear spin. The experiments are implemented on the two-qubit system composed of four energy levels $|m_S=0, m_I=+1\rangle$, $|m_S=-1, m_I=+1\rangle$, $|m_S=0, m_I=0\rangle$, and $|m_S=-1, m_I=0\rangle$ labeled by $|0\rangle_e |1\rangle_n$, $|0\rangle_e |0\rangle_n$, $|-1\rangle_e |1\rangle_n$, and $|-1\rangle_e |0\rangle_n$. The electronic zero-filed splitting is $D=2.87$ GHz and the Zeeman splitting of the electron spin is $\omega_e$. The two-qubit system is controlled by two microwave (MW) pulses (blue arrows) and two radio-frequency (RF) pulses (orange arrows), which selectively drive the two electron-spin transitions and the two nuclear-spin transitions, respectively. \textbf{c}, Quantum circuit of the experiment. The electron spin qubit is taken as the system qubit while the nuclear spin qubit is served as the ancilla qubit. X and Y denote the single nuclear spin qubit rotation around the \textit x and \textit y axes. The two-qubit system is prepared to $|\Psi(0)\rangle = |0\rangle_e |-\rangle_n + \eta_0(0)|0\rangle_e |+\rangle_n$ by rotations $\mathrm{Y}(\theta)$ and $\mathrm{X}(\pi/2)$. Then the two-qubit system evolve under the dilation Hamiltonian $H_{s,a}(t)$. The populations of the four energy levels are measured after the rotation $\mathrm{X}(-\pi/2)$.
 }
    \label{Fig1}
\end{figure}

A solid-state spin system based on NV center in diamond is utilized to demonstrate our proposal.
As depicted in Fig.~\ref{Fig1}a, the NV center consists of a substitutional nitrogen atom with an adjacent vacancy site in the diamond crystal lattice.
The Hamiltonian of the electron spin and the $^{14}$N nuclear spin system is
\begin{equation}
H_{\mathrm{NV}} = 2\pi(DS_z^2 + \omega_eS_z + QI_z^2 + \omega_nI_z + AS_zI_z),
\end{equation}
where $S_z$ and $I_z$ are the spin operators of the electron spin (spin-1) and the nuclear spin (spin-1).
The electronic zero-field splitting is $D=2.87$ GHz and the nuclear quadrupolar interaction is $Q=-4.95$ MHz.
The two spins are coupled by an hyperfine interaction $A=-2.16$ MHz.
A magnetic field is applied along the NV symmetry axis ([1 1 1] crystal axis) to remove the degeneracy of the $|m_S=\pm1\rangle$ states, yielding the electron and nuclear Zeeman frequencies $\omega_e$ and $\omega_n$, respectively.
A subspace of the total system is utilized to consist a two-qubit system, which is spanned by the four energy levels $|m_S=0, m_I=+1\rangle$, $|m_S=0, m_I=0\rangle$, $|m_S=-1, m_I=+1\rangle$, and $|m_S=-1, m_I=0\rangle$ labeled by $|0\rangle_e |1\rangle_n$, $|0\rangle_e |0\rangle_n$, $|-1\rangle_e |1\rangle_n$, and $|-1\rangle_e |0\rangle_n$ as shown in Fig.~\ref{Fig1}b.
The electron spin qubit is treated as the system qubit while the nuclear spin qubit is served as the ancilla qubit.
The dilated Hamiltonian $H_{s,a}(t)$ is achieved by two selective microwave (MW) pulses which are simultaneously applied on the electron spin.
The control Hamiltonian can be written as
\begin{equation}
\begin{aligned}
H_c &= 2\pi\sqrt{2}\Omega_1(t)\cos[\int_0^t\omega_1(\tau)d\tau+\phi_1(t)] S_x \otimes |1\rangle_{nn}\langle1|  \\
&+2\pi\sqrt{2}\Omega_2(t)\cos[\int_0^t\omega_2(\tau)d\tau+\phi_2(t)] S_x \otimes |0\rangle_{nn}\langle0|,
\end{aligned}
\end{equation}
where $\Omega_1(t)$, $\omega_1(t)$ and $\phi_1(t)$ ($\Omega_2(t)$, $\omega_2(t)$ and $\phi_2(t)$) correspond to the Rabi frequency, angular
frequency, and phase of the selective control pulses on the electron spin which drive the spin transition if the state of the nuclear spin is $|1\rangle_n$ ($|0\rangle_n$).
The frequencies of the two MW pulses are set to be $\omega_1(t)=\omega_{\mathrm{MW1}}+2A_4(t)$ and $\omega_2(t)=\omega_{\mathrm{MW2}}-2A_4(t)$, respectively.
In the interaction picture, the two-qubit subspace of the total Hamiltonian, $H_{\mathrm{NV}}+H_c$, can be written as (see Supplementary Material for details)
\begin{equation}
\begin{aligned}
H_{\mathrm{rot}} &= \pi\Omega(t)\cos[\phi(t)]\sigma_x\otimes I + A_2(t)I\otimes\sigma_z\\ &+ \pi\Omega(t)\sin[\phi(t)]\sigma_y\otimes\sigma_z +  A_4(t)\sigma_z\otimes\sigma_z
\end{aligned}
\end{equation}
by choosing $\Omega_1(t)=\Omega_2(t)=\Omega(t)$ and $-\phi_1(t)=\phi_2(t)=\phi(t)$.
The dilated Hamiltonian $H_{s,a}(t)$ can realized when $\Omega(t)=\sqrt{A_1^2(t)+A_3^2(t)}/\pi$ and $\phi(t)=\arctan(A_3(t)/A_1(t))$.

Our experiment was implemented on a NV center in $[100]$ face bulk diamond which was isotopically purified ([$^{12}$C]=99.9\%).
The dephasing time $T_2^*$ of the electron spin is 19 (2) $\mathrm{\mu s}$.
The 532 nm green laser pulses were modulated by an acousto-optic modulator (ISOMET).
The laser beam traveled twice through the acousto-optic modulator before going through an oil objective (Olympus, PLAPON 60*O, NA 1.42).
The phonon sideband fluorescence (wavelength, 650-800nm) went through the same oil objective and was collected by an avalanche photodiode (Perkin Elmer, SPCM-AQRH-14) with a counter card.
The magnetic field of 506 G was provided by a permanent magnet along the NV symmetry axis and the state of the two-qubit system can be effectively polarized to $|0\rangle_e|1\rangle_n$ by laser pumping.
An arbitrary waveform generator (Keysight M8190A) generated microwave and radio-frequency pulses to manipulate the states of the two-qubit system.
The microwave pulses were amplified by power amplifiers (Mini Circuits ZHL-30W-252-S+) and fed by a broadband coplanar waveguide with $15~$GHz
bandwidth.
The radio-frequency pulses were carried by a home-built coil with dual resonance frequencies after a power amplifier (Mini Circuits LZY-22+).

The experiment was preformed on a home-built optical detected magnetic resonance setup.
When the strength of the static magnetic field was set to 506 Gauss, optical pumping laser pulses polarized the electron spin and nuclear spin simultaneously into the state $|0\rangle_e |1\rangle_n$ owing to resonant polarization exchange with the electronic spin in the excited state\cite{PRL_Wrachtrup}.
The initial state of the two-qubit is $|\Psi\rangle = |0\rangle_e |-\rangle_n + \eta_0(0)|0\rangle_e |+\rangle_n$, which was obtained by the single-qubit rotation $Y(\theta)$ followed by the rotation $X(\pi/2)$ on nuclear spin as show in Fig.~\ref{Fig1}c.
The operator, $Y(\theta)$, stands for the rotation around \textit y axis with the rotation angle, $\theta = 2\arctan[\eta_0(0)]$.
The rotation angle, $\theta$, varies with different $\mathcal{PT}$ symmetric Hamiltonian $H_s$.
The rotation $X(\pi/2)$ is the single-qubit rotations around \textit x axis to realize transformation between the basis spanned by $\{|0\rangle_n, |1\rangle_n\}$ and the basis spanned by $\{|+\rangle_n, |-\rangle_n\}$ of the nuclear spin qubit.
Two selective MW pulses were applied on the electron spin to realize the dilation Hamiltonian $H_{s,a}(t)$.
The parameters $\Omega_1(t)$, $\Omega_2(t)$, $\omega_1(t)$, $\omega_2(t)$, $\phi_1(t)$ and $\phi_2(t)$ are chosen corresponding to $A_1(t)$, $A_2(t)$, $A_3(t)$ and $A_4(t)$ as mentioned above.
The MW pulses were generated by an arbitrary waveform generator and fed by a coplanar waveguide.
The nuclear spin rotation, $X(-\pi/2)$, transform the state $|\Psi(t)\rangle = |\psi(t)\rangle_e |-\rangle_n + \eta_0(t)|\psi(t)\rangle_e |+\rangle_n$ into $|\Phi(t)\rangle = |\psi(t)\rangle_e |1\rangle_n + \eta_0(t)|\psi(t)\rangle_e |0\rangle_n$.
Then the populations of each energy level of the two-qubit system are detected (see Supplementary Material for details).
All the single nuclear spin rotations were realized with two channel radio frequency (RF) pulses applied simultaneously on the nuclear spin as illustrated in Fig.~\ref{Fig1}b.
The frequency of the RF pulses are 2.9 MHz and 5.1 MHz corresponded to the nuclear spin transitions.
The RF pulses were fed on the nuclear spin by a home-built coil with dual resonance frequencies.
The Rabi frequency of the RF pulses were calibrated to 25 kHz.

\begin{figure}
\centering
\includegraphics[width=0.9\columnwidth]{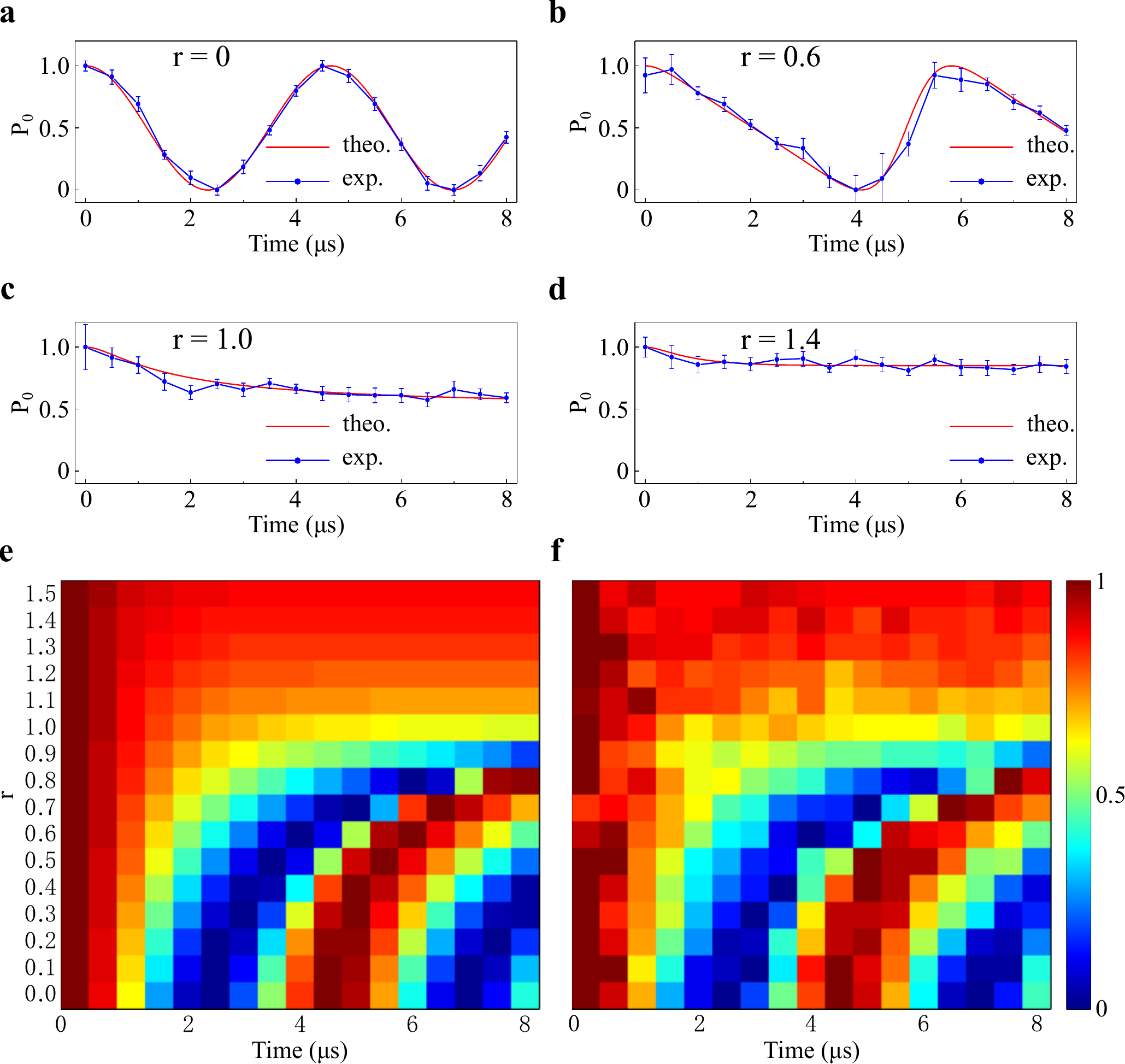}
    \caption{State evolution under $\mathcal{PT}$ symmetric Hamiltonian. \textbf{a}-\textbf{d}, Experimental state evolution under $\mathcal{PT}$ symmetric  Hamiltonians with different parameter a of the Hamiltonian, where $r=0$ (\textbf{a}), $r=0.6$ (\textbf{b}), $r=1.0$ (\textbf{c}) and $r=1.4$ (\textbf{d}) correspond to the Hermitian, $\mathcal{PT}$ unbroken, exceptional point and $\mathcal{PT}$ broken case, respectively. $P_0$ is the renormalized population of the state $|m_S=0\rangle$ of the electron spin when the nuclear spin state is in the selected state $|m_I=+1\rangle$. Blue dots are experimental results, and red lines are the theoretical predictions. \textbf{e} (theoretical results ) and \textbf{f} (experimental results) plot results for various $r$ values. The color bar stands for the population $P_0$.  }
    \label{Fig2}
\end{figure}

The state evolution under the $\mathcal{PT}$ symmetric Hamiltonian $H_s$ is explored by monitoring $P_0$, i.e., the renormalized population of the state $|m_S=0\rangle$ of the electron spin when the nuclear spin state is in the selected state $|m_I=+1\rangle$ (see supplementary materials for details).
The time of the state evolution is varied from 0 to 8 $\mathrm{\mu s}$.
Figure \ref{Fig2}a-d show the state evolution under $\mathcal{PT}$ symmetric Hamiltonian with the Hermitian case $r=0$ ( Fig.~\ref{Fig2}a ), $\mathcal{PT}$ unbroken case $r=0.6$ ( Fig.~\ref{Fig2}b ), exceptional point case $r=1.0$ ( Fig.~\ref{Fig2}c ) and $\mathcal{PT}$ broken case $r=1.4$ ( Fig.~\ref{Fig2}d ).
All errors are one standard deviation with repeating the experiments for 0.5 million times.
In the $\mathcal{PT}$ unbroken region, the time evolution of the state is periodic oscillation, while the oscillation breaks down in the $\mathcal{PT}$ broken region.
The state evolutions under $\mathcal{PT}$ symmetric Hamiltonian with various values of the parameter, $r$, are shown in Fig.~\ref{Fig2}e-f.
The $\mathcal{PT}$ phase transition threshold is evidently revealed at the exceptional point $r=1.0$.
The experimental results (Fig.~\ref{Fig2}f) show good agreement with the corresponding theoretical predictions (Fig.~\ref{Fig2}e).

\begin{figure}
\centering
\includegraphics[width=0.9\columnwidth]{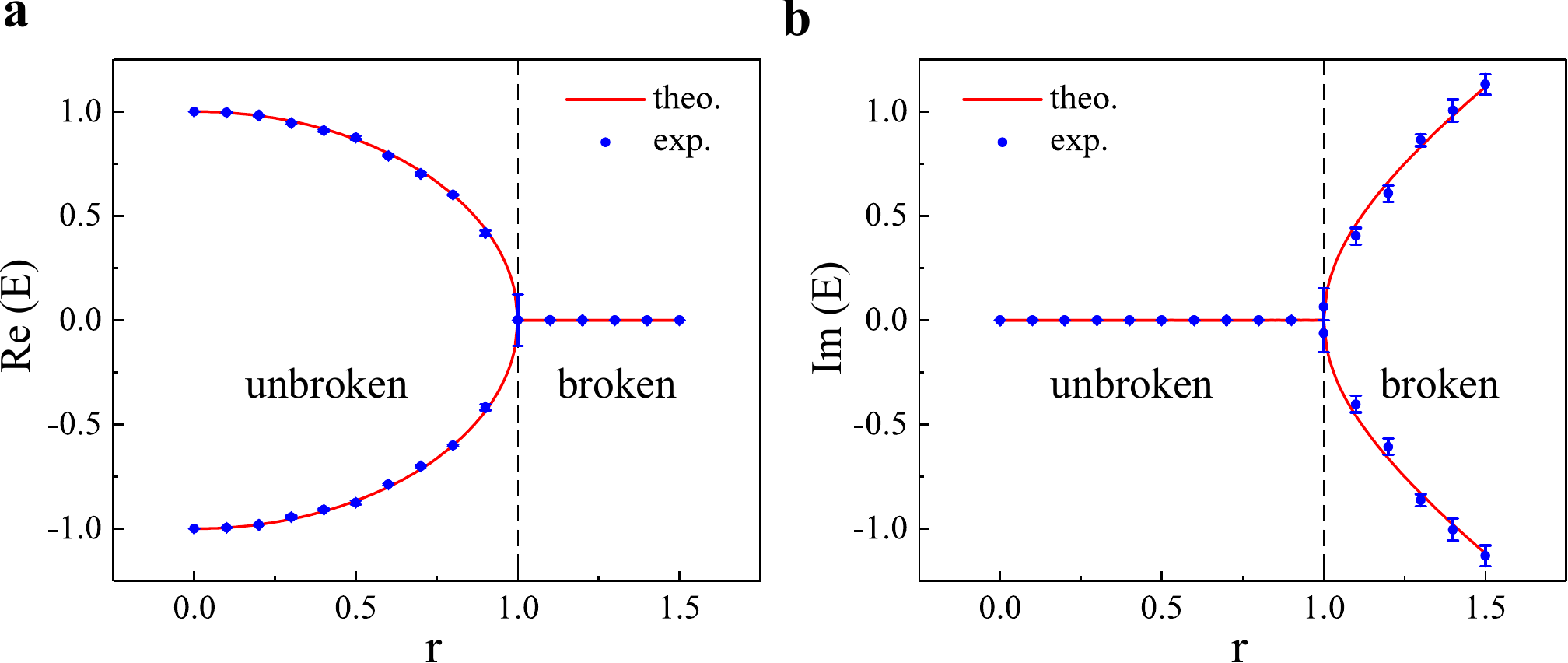}
    \caption{Experimental observation the breaking of the $\mathcal{PT}$ symmetry. \textbf{a} Real part and \textbf{b} imaginary part of the eigenvalues of the $\mathcal{PT}$ symmetric Hamiltonian. The blue dots are experimental data, and the red lines are the theoretical predictions of the eigenvalues. The  $0<r<1$ regime represents the $\mathcal{PT}$ unbroken case and the $r>1$ regime represents the $\mathcal{PT}$ broken case. The exceptional point occurs at $r=1$.
 }
    \label{Fig3}
\end{figure}

The $\mathcal{PT}$ phase transition is also characterized by the eigenvalues of the $\mathcal{PT}$ symmetric Hamiltonians as shown in Fig.~\ref{Fig3}.
The eigenvalues of the $H_s$ can be achieved by $E_\pm=\pm\sqrt{1-r_{\mathrm{exp}}^2}$.
The parameter $r_{\mathrm{exp}}$ is obtained by curve fitting the experimental time evolution of the population $P_0$ to theoretical predictions under $\mathcal{PT}$ symmetric Hamiltonian $H_s$ (see supplementary materials for the details).
When $r<1$, the system is in a $\mathcal{PT}$ symmetry unbroken region. The eigenvalues $E$ are still real as $r$ approaches 1 from 0.
At the exceptional point $r=1$, the eigenvalues $E$ coalesce to 0.
The system is in the $\mathcal{PT}$ symmetry broken region when $r>1$.
The real parts of the eigenvalues $E$ coalesce and the imaginary parts appears.
The experimental results show excellent agreement with the corresponding theoretical predictions.

In summary, we have experimentally demonstrate the state evolution under $\mathcal{PT}$ symmetric Hamiltonian by the Hermitian dilation method.
The  breaking of $\mathcal{PT}$ symmetry has been observed in a single electron spin.
The universal dilation method present here is compatible for Hermitian dilation of an arbitrary non-Hermitian Hamiltonian,  thus our work opens a door for future experimental study of the intriguing non-Hermitian and $\mathcal{PT}$ symmetric physics with quantum systems.

This work was supported by the National Key R$\&$D Program of China (Grants No. 2018YFA0306600 and No. 2016YFB0501603), the CAS (Grants No. GJJSTD20170001, No.QYZDY-SSW-SLH004 and No.QYZDB-SSW-SLH005), and Anhui Initiative in Quantum Information Technologies (Grant No.
AHY050000). X.R. thanks the Youth Innovation Promotion Association of Chinese Academy of Sciences for the support.

\onecolumngrid
\vspace{1.5cm}
\begin{center}
\textbf{\large APPENDIX}
\end{center}

\setcounter{figure}{0}
\setcounter{equation}{0}
\setcounter{table}{0}
\makeatletter
\renewcommand{\thefigure}{S\arabic{figure}}
\renewcommand{\theequation}{S\arabic{equation}}
\renewcommand{\thetable}{S\arabic{table}}
\renewcommand{\bibnumfmt}[1]{[RefS#1]}
\renewcommand{\citenumfont}[1]{RefS#1}



\section{\uppercase\expandafter{\romannumeral1}. Universal Hermitian dilation of non-Hermitian Hamiltonians }

This section mainly focus on universally dilating a non-Hermitian Hamiltonian into a Hermitian one.
Subsection A demonstrates the dilation method, followed by some derivation used in the dilation in subsection B and subsection C.
Finally, in subsection D the universality of this method is proved.

\subsection{A. Universal dilation method }

Our target is to realize the dynamics of a quantum system $s$, which is described by the evolution $\varepsilon_1(t)$ governed by Hamiltonian $H_s(t)$.
$H_s(t)$ is non-Hermitian, arbitrary dimensional and time-dependent in general.
The state evolution of system $s$ is described by $|\psi(t)\rangle$, so the corresponding Schr\"{o}dinger equation can be written as (natural unit are chosen so that $\hbar=1$ in this supplementary material)
\begin{equation}
\label{Eq1}
i\frac{d}{dt}|\psi(t)\rangle=H_s(t)|\psi(t)\rangle.
\end{equation}

To realize $\varepsilon_1(t)$ in a quantum system, an ancilla qubit is introduced to dilate $H_s(t)$ into a Hermitian Hamiltonian $H_{s,a}(t)$. The evolution of the combined system under $H_{s,a}(t)$ is $\varepsilon_2(t)$.
The state evolution of the combined system is described by $|\Psi(t)\rangle$, which satisfies the following Schr\"{o}dinger equation
\begin{equation}
\label{Eq2}
i\frac{d}{dt}|\Psi(t)\rangle=H_{s,a}(t)|\Psi(t)\rangle.
\end{equation}
$|\Psi(t)\rangle$ is a dilation of $|\psi(t)\rangle$ and can be written as
\begin{equation}
\label{Eq3}
\left\{
\begin{aligned}
& |\Psi(t)\rangle=|\psi(t)\rangle|-\rangle+|\chi(t)\rangle|+\rangle,  \\
& |\chi(t)\rangle=\eta(t)|\psi(t)\rangle,  \\
\end{aligned}
\right.
\end{equation}
where $|-\rangle=\frac{|0\rangle-i|1\rangle}{\sqrt{2}}$ and $|+\rangle=-i\frac{|0\rangle+i|1\rangle}{\sqrt{2}}$ are the eigenstates of Pauli operator $\sigma_y$, which forms an orthonormal basis of the ancilla qubit, and $\eta(t)$ is a linear operator.
When a measurement is applied on the ancilla qubit and $|-\rangle$ is postselected, the evolution $\varepsilon_1(t)$ governed by the non-Hermitian Hamiltonian $H_s(t)$ can be produced.

Now the key is to derive the expression of $H_{s,a}(t)$.
The mathematical form of $H_{s,a}(t)$ can be written as
\begin{equation}
\label{Eq4}
H_{s,a}(t)=H_{s,a}^{(++)}(t)\otimes|+\rangle\langle+|+H_{s,a}^{(+-)}(t)\otimes|+\rangle\langle-|+H_{s,a}^{(-+)}(t)\otimes|-\rangle\langle+|+H_{s,a}^{(--)}(t)\otimes|-\rangle\langle-|,
\end{equation}
where $H_{s,a}^{(++)}(t),H_{s,a}^{(+-)}(t),H_{s,a}^{(-+)}(t)$ and $H_{s,a}^{(--)}(t)$ have the same dimension as $H_s(t)$.
Due to the Hermiticity of $H_{s,a}(t)$, we have
\begin{equation}
\label{Eq5}
\left\{
\begin{aligned}
&H_{s,a}^{(++)\dagger}(t)=H_{s,a}^{(++)}(t),  \\
&H_{s,a}^{(+-)\dagger}(t)=H_{s,a}^{(-+)}(t),   \\
&H_{s,a}^{(--)\dagger}(t)=H_{s,a}^{(--)}(t).   \\
\end{aligned}
\right.
\end{equation}
By substituting equations \ref{Eq1}, \ref{Eq3} and \ref{Eq4} into equation \ref{Eq2} and rearranging terms, we obtain
\begin{equation}
\label{Eq6}
\left\{
\begin{aligned}
&H_{s,a}^{(--)}(t)+H_{s,a}^{(-+)}(t)\eta(t)-H_s(t)=0,    \\
&H_{s,a}^{(+-)}(t)+H_{s,a}^{(++)}(t)\eta(t)-i\frac{d}{dt}\eta(t)-\eta(t)H_s(t)=0.   \\
\end{aligned}
\right.
\end{equation}
After taking Hermitian transpose of operators in equation \ref{Eq6} we can get
\begin{equation}
\label{Eq7}
\left\{
\begin{aligned}
&  H_{s,a}^{(--)}(t)+\eta^\dag(t)H_{s,a}^{(+-)}(t)-H_s^\dag(t)=0,   \\
&  H_{s,a}^{(-+)}(t)+\eta^\dag(t)H_{s,a}^{(++)}(t)+i\frac{d}{dt}\eta^\dag(t)-H_s^\dag(t)\eta^\dag(t)=0,   \\
\end{aligned}
\right.
\end{equation}
where equation \ref{Eq5} has been taken into consideration.
Finally, equations \ref{Eq6} and \ref{Eq7} reduce to
\begin{equation}
\label{Eq8}
\left\{
\begin{aligned}
&H_{s,a}^{(--)}(t)=H_s(t)-[H_s^\dag(t)\eta^\dag(t)-i\frac{d}{dt}\eta^\dag(t)-\eta^\dag(t)H_{s,a}^{(++)}(t)]\eta(t),   \\
&H_{s,a}^{(-+)}(t)=H_s^\dag(t)\eta^\dag(t)-i\frac{d}{dt}\eta^\dag(t)-\eta^\dag(t)H_{s,a}^{(++)}(t),  \\
&H_{s,a}^{(+-)}(t)=\eta(t)H_s(t)+i\frac{d}{dt}\eta(t)-H_{s,a}^{(++)}(t)\eta(t), \\
\end{aligned}
\right.
\end{equation}
together with an equation that $\eta(t)$ should satisfy,
\begin{equation}
\label{Eq9}
i\frac{d}{dt}[\eta^\dag(t)\eta(t)]=H_s^\dag(t)[\eta^\dag(t)\eta(t)+I]-[\eta^\dag(t)\eta(t)+I]H_s(t).
\end{equation}
To solve equation \ref{Eq9}£©, we can define a Hermitian operator $M(t)$ as follows
\begin{equation}
\label{Eq10}
M(t)\equiv\eta^\dag(t)\eta(t)+I,
\end{equation}
where $I$ is the identity operator.
Then equation \ref{Eq9} can be rewritten as
\begin{equation}
\label{Eq11}
i\frac{d}{dt}M(t)=H_s^\dag(t)M(t)-M(t)H_s(t).
\end{equation}
The solution of equation \ref{Eq11} takes the form
\begin{equation}
\label{Eq12}
M(t)=\mathcal{T}e^{-i\int_{0}^{t}H_s^\dag(t)dt}M(0)\overline{\mathcal{T}}e^{i\int_{0}^{t}H_s(t)dt},
\end{equation}
where $\mathcal{T}$ and $\overline{\mathcal{T}}$ are time-ordering and anti-time-ordering operators, respectively. $M(0)$ is an initial operator of operator $M(t)$, which is chosen to ensure that $M(t)-I$ keeps positive for all $t$.
According to equation \ref{Eq10}, an expression of $\eta(t)$ can be written as
\begin{equation}
\label{Eq13}
\eta(t)=U(t)[M(t)-I]^\frac{1}{2},
\end{equation}
where $U(t)$ is an arbitrary differentiable unitary operator.
Equation \ref{Eq4}, \ref{Eq8}, \ref{Eq12} together with equation \ref{Eq13} give an explicit expression of the dilated Hamiltonian $H_{s,a}(t)$, in which $H_{s,a}^{(++)}(t)$ is an arbitrary Hamiltonian operator.

Note that $H_{s,a}(t)$ is not uniquely determined since $H_{s,a}^{(++)}(t)$ can be an arbitrary Hermitian operator and $U(t)$ can be an arbitrary differentiable unitary operator.
This arbitrariness makes it possible to flexible design $H_{s,a}(t)$ according to different experimental systems.
For example, to make $H_{s,a}(t)$ easy to be constructed in NV center, we can choose
\begin{equation}
\label{Eq14}
\left\{
\begin{aligned}
& U(t)=I,  \\
& H_{s,a}^{(++)}(t)=\{H_s(t)+[i\frac{d}{dt}\eta(t)+\eta(t)H_s(t)]\eta(t)\}M^{-1}(t).
\end{aligned}
\right.
\end{equation}
The Hermiticity of $H_{a,s}^{(++)}(t)$ will be proved in subsection B of this section. According to equation \ref{Eq14}, equation \ref{Eq13} reduces to
\begin{equation}
\label{Eq15}
\eta(t)=[M(t)-I]^\frac{1}{2},
\end{equation}
which means $\eta(t)$ is also a Hermitian operator.
In this case the Hermitian operator,
\begin{equation}
\label{Eq16}
M(t)=\eta^\dag(t)\eta(t)+I=\eta^2(t)+I,
\end{equation}
commutes with $\eta(t)$, and the inverse operator of $M(t)$, $M^{-1}(t)$, is Hermitian and commutes with $\eta(t)$ as well.
By substituting equation \ref{Eq14} into equation \ref{Eq8} and considering the Hermiticity of $\eta(t)$, we have
\begin{equation}
\label{Eq17}
\left\{
\begin{aligned}
&H_{s,a}^{(--)}(t)=H_s(t)-H_s^\dag(t)\eta^2(t)+i[\frac{d}{dt}\eta(t)]\eta(t)+\eta(t)\{H_s(t)+[i\frac{d}{dt}\eta(t)+\eta(t)H_s(t)]\eta(t)\}M^{-1}(t)\eta(t),    \\
&H_{s,a}^{(-+)}(t)=H_s^\dag(t)\eta(t)-i\frac{d}{dt}\eta(t)-\eta(t)\{H_s(t)+[i\frac{d}{dt}\eta(t)+\eta(t)H_s(t)]\eta(t)\}M^{-1}(t),                         \\
&H_{s,a}^{(+-)}(t)=\eta(t)H_s(t)+i\frac{d}{dt}\eta(t)-\{H_s(t)+[i\frac{d}{dt}\eta(t)+\eta(t)H_s(t)]\eta(t)\}M^{-1}(t)\eta(t).        \\
\end{aligned}
\right.
\end{equation}
Equation \ref{Eq17} can be simplified to
\begin{equation}
\label{Eq18}
\left\{
\begin{aligned}
&H_{s,a}^{(--)}(t)=\{H_s(t)+[i\frac{d}{dt}\eta(t)+\eta(t)H_s(t)]\eta(t)\}M^{-1}(t),    \\
&H_{s,a}^{(-+)}(t)=[H_s(t)\eta(t)-\eta(t)H_s(t)-i\frac{d}{dt}\eta(t)]M^{-1}(t),           \\
&H_{s,a}^{(+-)}(t)=-[H_s(t)\eta(t)-\eta(t)H_s(t)-i\frac{d}{dt}\eta(t)]M^{-1}(t).          \\
\end{aligned}
\right.
\end{equation}
The detail of the derivation from equation \ref{Eq17} to equation \ref{Eq18} is given in subsection C of this section.
Substituting equation \ref{Eq18} and equation \ref{Eq14} into equation \ref{Eq4}, we obtain
\begin{equation}
\label{Eq19}
\begin{aligned}
H_{s,a}(t)=
&\{\{H_s(t)+[i\frac{d}{dt}\eta(t)+\eta(t)H_s(t)]\eta(t)\}\otimes|+\rangle\langle+|-[H_s(t)\eta(t)-\eta(t)H_s(t)-i\frac{d}{dt}\eta(t)]\otimes|+\rangle\langle-|    \\
&+[H_s(t)\eta(t)-\eta(t)H_s(t)-i\frac{d}{dt}\eta(t)]\otimes|-\rangle\langle+|+\{H_s(t)+[i\frac{d}{dt}\eta(t)+\eta(t)H_s(t)]\eta(t)\}\otimes|-\rangle\langle-|\}M^{-1}(t),
\end{aligned}
\end{equation}
or
\begin{equation}
\label{Eq20}
H_{s,a}(t)=\Lambda(t)\otimes I+\Gamma(t)\otimes\sigma_z,
\end{equation}
with
\begin{equation}
\label{Eq21}
\left\{
\begin{aligned}
&\Lambda(t)=\{H_s(t)+[i\frac{d}{dt}\eta(t)+\eta(t)H_s(t)]\eta(t)\}M^{-1}(t),    \\
&\Gamma(t)=i[H_s(t)\eta(t)-\eta(t)H_s(t)-i\frac{d}{dt}\eta(t)]M^{-1}(t).           \\
\end{aligned}
\right.
\end{equation}

\subsection{B. Proof of the Hermiticity of $H_{s,a}^{(++)}(t)$}

Utilizing the commutation relation between $M(t)$ and $\eta(t)$, the second formula of equation \ref{Eq14} can be rewritten as
\begin{equation}
\label{Eq22}
H_{s,a}^{(++)}(t)=M^{-1}(t)\{M(t)H_s(t)+\eta(t)M(t)H_s(t)\eta(t)+iM(t)[\frac{d}{dt}\eta(t)]\eta(t)\}M^{-1}(t).
\end{equation}
By substituting equation \ref{Eq16} into the third term of the right-hand side of equation \ref{Eq22}, we obtain
\begin{equation}
\label{Eq23}
H_{s,a}^{++}(t)=M^{-1}(t)\{M(t)H_s(t)+\eta(t)M(t)H_s(t)\eta(t)+i\eta^2(t)[\frac{d}{dt}\eta(t)]\eta(t)+i[\frac{d}{dt}\eta(t)]\eta(t)\}M^{-1}(t).
\end{equation}
Considering the Hermiticity of $\eta(t)$, $M(t)$ and $M^{-1}(t)$, the Hermitian conjunction operator of $H_{a,s}^{(++)}(t)$ can be written as
\begin{equation}
\label{Eq24}
H_{s,a}^{(++)\dag}(t)=M^{-1}(t)\{H_s^\dag(t)M(t)+\eta(t)H_s^\dag(t)M(t)\eta(t)-i\eta(t)[\frac{d}{dt}\eta(t)]\eta^2(t)-i\eta(t)[\frac{d}{dt}\eta(t)]\}M^{-1}(t).
\end{equation}
Combining equation \ref{Eq23} and equation \ref{Eq24}, we have
\begin{equation}
\label{Eq25}
\begin{aligned}	
&H_{s,a}^{(++)\dag}(t)-H_{s,a}^{++}(t)     \\
=&M^{-1}(t)[H_s^\dag(t)M(t)-M(t)H_s(t)]M^{-1}(t)+M^{-1}(t)\eta(t)[H_s^\dag(t)M(t)-M(t)H_s(t)]\eta(t)M^{-1}(t) \\
&-iM^{-1}(t)\eta(t)\{[\frac{d}{dt}\eta(t)]\eta(t)+\eta(t)[\frac{d}{dt}\eta(t)]\}\eta(t)M^{-1}(t)-iM^{-1}(t)\{\eta(t)[\frac{d}{dt}\eta(t)]+\frac{d}{dt}\eta(t)]\eta(t)\}M^{-1}(t).
\end{aligned}	
\end{equation}
According to equation \ref{Eq11} and equation \ref{Eq16}, we have
\begin{equation}
\label{Eq26}
i\frac{d}{dt}M(t)=H_s^\dag(t)M(t)-M(t)H_s(t)=i\{[\frac{d}{dt}\eta(t)]\eta(t)+\eta(t)[\frac{d}{dt}\eta(t)]\}.
\end{equation}
Then, equation \ref{Eq25} vanishes, so the Hermiticity of $H_{s,a}^{(++)}(t)$ is proved.

\subsection{C. Derivation of the form of $H_{s,a}^{(--)}(t), H_{s,a}^{(-+)}(t)$ and $H_{s,a}^{(+-)}(t)$}

The first formula in equation \ref{Eq17} reads
\begin{equation}
\label{Eq27}
H_{s,a}^{(--)}(t)=H_s(t)-H_s^\dag(t)\eta^2(t)+i[\frac{d}{dt}\eta(t)]\eta(t)+\eta(t)\{H_s(t)+[i\frac{d}{dt}\eta(t)+\eta(t)H_s(t)]\eta(t)\}M^{-1}(t)\eta(t).
\end{equation}
Substitute equation \ref{Eq16} into equation \ref{Eq27} and consider the commutation of $M^{-1}(t)$ and $\eta(t)$  we have
\begin{equation}
\label{Eq28}
\begin{aligned}
H_{s,a}^{(--)}(t)=&-H_s^\dag(t)M(t)+H_s^\dag(t)+i[\frac{d}{dt}\eta(t)]\eta(t)+\eta(t)H_s(t)\eta(t)M^{-1}(t)\\ &+i\eta(t)[\frac{d}{dt}\eta(t)][M(t)-I]M^{-1}(t)+[M(t)-I]H_s(t)[M(t)-I]M^{-1}(t)+H_s(t).
\end{aligned}
\end{equation}
Simplify and rearrange terms in equation \ref{Eq28}, we obtain
\begin{equation}
\label{Eq29}
\begin{aligned}
H_{s,a}^{(--)}(t)=&-[H_s^\dag(t)M(t)-M(t)H_s(t)]+i\{[\frac{d}{dt}\eta(t)]\eta(t)+\eta(t)[\frac{d}{dt}\eta(t)]\}  \\  &  +\{H_s(t)+\eta(t)H_s(t)\eta(t)-i\eta(t)[\frac{d}{dt}\eta(t)]+[H_s^\dag(t)M(t)-M(t)H_s(t)]\}M^{-1}(t),
\end{aligned}
\end{equation}
then, by substituting equations \ref{Eq26} into equation \ref{Eq29} and rearranging terms, equation \ref{Eq29} reduces to
\begin{equation}
\label{Eq30}
H_{s,a}^{(--)}(t)=\{H_s(t)+[i\frac{d}{dt}\eta(t)+\eta(t)H_s(t)]\eta(t)\}M^{-1}(t),
\end{equation}
which is the first formula in equation \ref{Eq18}.

The third formula in equation \ref{Eq17} is given by
\begin{equation}
\label{Eq31}
H_{s,a}^{(+-)}(t)=\eta(t)H_s(t)+i\frac{d}{dt}\eta(t)-\{H_s(t)+[i\frac{d}{dt}\eta(t)+\eta(t)H_s(t)]\eta(t)\}M^{-1}(t)\eta(t).
\end{equation}
Considering the commutation of $M^{-1}(t)$ and $\eta(t)$, and substituting equation \ref{Eq16}, we have
\begin{equation}
\label{Eq32}
H_{s,a}^{(+-)}(t)=\eta(t)H_s(t)+i\frac{d}{dt}\eta(t)-H_s(t)\eta(t)M^{-1}(t)-[i\frac{d}{dt}\eta(t)+\eta(t)H_s(t)][M(t)-I]M^{-1}(t).
\end{equation}
Equation \ref{Eq32} reduces to
\begin{equation}
\label{Eq33}
H_{s,a}^{(+-)}(t)=-[H_s(t)\eta(t)-\eta(t)H_s(t)-i\frac{d}{dt}\eta(t)]M^{-1}(t),
\end{equation}
so the third formula in equation \ref{Eq18} is proved.

To prove the second formula in equation \ref{Eq18}, we rewrite equation \ref{Eq33} as
\begin{equation}
\label{Eq34}
H_{s,a}^{(+-)}(t)=-M^{-1}(t)[M(t)H_s(t)\eta(t)-\eta(t)M(t)H_s(t)-i\eta^2(t)\frac{d}{dt}\eta(t)-i\frac{d}{dt}\eta(t)]M^{-1}(t),
\end{equation}
where equation \ref{Eq16} and commutation of $M(t)$ and $\eta(t)$ have been utilized.
According to equation \ref{Eq5}, $H_{s,a}^{(-+)}(t)=H_{s,a}^{(+-)\dag}(t)$, we have
\begin{equation}
\label{Eq35}
H_{s,a}^{(-+)}(t)=-M^{-1}(t)\{\eta(t)H_s^\dag(t)M(t)-H_s^\dag(t)M(t)\eta(t)+i[\frac{d}{dt}\eta(t)]\eta^2(t)+i\frac{d}{dt}\eta(t)\}M^{-1}(t),
\end{equation}
where the Hermiticity of $\eta(t)$, $M(t)$ and $M^{-1}(t)$ have been considered.
Combining equation \ref{Eq34} and equation \ref{Eq35}, we obtain
\begin{equation}
\label{Eq36}
\begin{aligned}
H_{s,a}^{(-+)}(t)+H_{s,a}^{(+-)}(t)=& -M^{-1}(t)\{\eta(t)[H_s^\dag(t)M(t)-M(t)H_s(t)]- \\ &[H_s^\dag(t)M(t)-M(t)H_s(t)]\eta(t)+i[\frac{d}{dt}\eta(t)]\eta^2(t)-i\eta^2(t)\frac{d}{dt}\eta(t)\}M^{-1}(t).
\end{aligned}
\end{equation}
Substituting equation \ref{Eq26} into equation \ref{Eq36}, we have
 \begin{equation}
\label{Eq37}
\begin{aligned}
H_{s,a}^{(-+)}(t)+H_{s,a}^{(+-)}(t)=
&-M^{-1}(t)\{i\eta(t)[\frac{d}{dt}\eta(t)]\eta(t)+i\eta^2(t)[\frac{d}{dt}\eta(t)]-i[\frac{d}{dt}\eta(t)]\eta^2(t)-i\eta(t)[\frac{d}{dt}\eta(t)]\eta(t)   \\
&+i[\frac{d}{dt}\eta(t)]\eta^2(t)-i\eta^2(t)\frac{d}{dt}\eta(t)\}M^{-1}(t).
\end{aligned}
\end{equation}
By canceling terms, equation \ref{Eq37} vanishes, that is
\begin{equation}
\label{Eq38}
H_{s,a}^{(-+)}(t)=-H_{s,a}^{(+-)}(t)=[H_s(t)\eta(t)-\eta(t)H_s(t)-i\frac{d}{dt}\eta(t)]M^{-1}(t),
\end{equation}
which is the second formula in equation \ref{Eq18}.

\subsection{D. Proof of the universality of this dilation method}

Here we show that our method can be utilized for Hermitian dilation of an arbitraty Hamiltonian $H_s(t)$.
Without loss of generality, the evolution time is denoted by $T$, so $t\in[0,T]$ during the evolution.
Since $H_s(t)$ is a Hamiltonian governing the evolution of a quantum system, given the state $|\psi(t_1)\rangle=|\psi_1\rangle$ at any time point $t_1\in[0,T]$, the subsequent state evolution is uniquely determined by $H_s(t)$.
That is, the Schr\"{o}dinger equation with an initial value,
\begin{equation}
\label{Eq39}
\left\{
\begin{aligned}
& i\frac{d}{dt}|\psi(t)\rangle=H_s(t)|\psi(t)\rangle,  \\
&|\psi(t_1)\rangle=|\psi_1\rangle,
\end{aligned}
\right.
\end{equation}
has a unique solution on $[t_1,T]$.
Stated another way, the evolution operator
\begin{equation}
\label{Eq40}
\varepsilon_1(t;t_1)=\mathcal{T}e^{-i\int_{t_1}^{t}H_s(t)dt},
\end{equation}
exist on $[t_1,T]$.
Obviously, $\varepsilon_1(t_1;t_1)=I$ for any $t_1\in[0,T]$.

It can be proved that $\varepsilon_1(t;t_1)$ is an invertible operator for $t\in[0,T]$ as follows.
Suppose that $\varepsilon_1(t;t_1)$ is not  invertible, then we can divide the evolution time into several segments, for example, $[t_1,t]=[t_1,t_2]\cup[t_2,t_3]\cup...\cup[t_{k-1},t_k]\cup[t_k,t]$, where $t_1<t_2<t_3<...<t_k<t$, in this case, we have
\begin{equation}
\label{Eq41}
\varepsilon_1(t;t_1)=\varepsilon_1(t;t_k)\varepsilon_1(t_k;t_{k-1})...\varepsilon_1(t_3;t_2)\varepsilon_1(t_2;t_1).
\end{equation}
If $\varepsilon_1(t;t_1)$ is not invertible, there must be at least a segment corresponding to which the evolution operator is not invertible.
Without loss of generality, suppose $\varepsilon_1(t_{n+1};t_n)$, the evolution operator on $[t_n,t_{n+1}]$ where $t_1<t_n<t_{n+1}<t$, is not invertible.
The segment $[t_n,t_{n+1}]$ can be further divided into several subsegments.
Similarity, there is at least a subsegment corresponding to which the evolution operator is not invertible.
The division can be implemented for arbitrary times.
As a result, there is a segment with infinitesimal length and the corresponding evolution operator is not invertible.
However, an evolution operator with infinitesimal time duration tends to $I$, which is invertible.
The contradiction shows that $\varepsilon_1(t;t_1)$ must be an invertible operator,then the inverse operator of $\varepsilon_1(t;t_1)$ can be written as
\begin{equation}
\label{Eq42}
\varepsilon_1^{-1}(t;t_1)=\overline{\mathcal{T}}e^{i\int_{t_1}^{t}H_s(t)dt}.
\end{equation}

According to the derivation of $H_{s,a}(t)$ in part one, the dilated Hamiltonian $H_{s,a}(t)$ can be obtained once a differentiable $\eta(t)$, which is a solution of equation \ref{Eq9}, is derived.
Equations \ref{Eq12} and \ref{Eq13} provide the solution to equation \ref{Eq9}, given that $M(t)$ is a positive operator with all eigenvalues larger than $1$ during the evolution.
It will be shown in the following that, by appropriate selection of the initial operator $M(0)$, the aforementioned requirement can be always satisfied for an arbitrary Hamiltonian $H_s(t)$.
To start with, select a positive operator $M_0^\prime$, of which all the eigenvalues are larger than $1$.
Then $M(0)^\prime$ is invertible and can be expressed as
\begin{equation}
\label{Eq43}
M_0^\prime=\zeta^\dag\zeta.
\end{equation}
Define
\begin{equation}
\label{Eq44}
M^\prime(t)\equiv\mathcal{T}e^{-i\int_{0}^{t}H_s^\dag(t)dt}M(0)^\prime\overline{\mathcal{T}}e^{i\int_{0}^{t}H_s(t)dt}
\end{equation}
then
\begin{equation}
\label{Eq45}
M^\prime(t)=(\zeta\overline{\mathcal{T}}e^{i\int_{0}^{t}H_s(t)dt})^\dag(\zeta\overline{\mathcal{T}}e^{i\int_{0}^{t}H_s(t)dt}),
\end{equation}
is a positive operator.
Because $\overline{\mathcal{T}}e^{i\int_{0}^{t}H_s(t)dt}=\varepsilon_1^{-1}(t;0)$, $M_0^\prime$, and $\mathcal{T}e^{-i\int_{0}^{t}H_s^\dag(t)dt}=[\varepsilon_1^{-1}(t;0)]^\dag$ are all invertible operators, then according to equation \ref{Eq45}, $M^\prime(t)$ is an invertible operator.
Therefore, all the eigenvalues of $M^\prime(t)$ are larger than $0$.
Suppose the minimum of the eigenvalues of $M^\prime(t)$ is $\mu^\prime$, then $\mu^\prime>0$.
Take $0<\mu<\mu^\prime$ and $M(0)=M_0^\prime/\mu$, then
\begin{equation}
\label{Eq46}
M(t)=\frac{1}{\mu}M^\prime(t),
\end{equation}
is a positive operator with all eigenvalues larger than $1$ during the evolution. Therefore, the dilated Hamiltonian $H_{s,a}(t)$ can be obtained with our method for an arbitrary Hamiltonian $H_s(t)$.


\section{\uppercase\expandafter{\romannumeral2}. Construct Hamiltonian $H_{s,a}(t)$ in NV center}

The dilated Hamiltonian shown in equation \ref{Eq20} takes the form
\begin{equation}
\label{Eq47}
H_{s,a}(t)=\Lambda(t)\otimes I+\Gamma(t)\otimes\sigma_z,
\end{equation}
where operators $\Lambda(t)$ and $\Gamma(t)$ are given in equation \ref{Eq21}.
By expanding $\Lambda(t)$ and $\Gamma(t)$ in terms of Pauli operators, we can rewrite $H_{s,a}(t)$ as
\begin{equation}
\label{Eq48}
\begin{aligned}
H_{s,a}(t)=
&B_1(t)I\otimes I+A_1(t)\sigma_x\otimes I + B_2(t)\sigma_y\otimes I+B_3(t)\sigma_z\otimes I+            \\
&A_2(t)I\otimes\sigma_z+B_4(t)\sigma_x\otimes\sigma_z +A_3(t)\sigma_y\otimes\sigma_z+A_4(t)\sigma_z\otimes\sigma_z,
\end{aligned}
\end{equation}
where $A_i(t)$ and $B_i(t)$, $i\in[1,4]$, are the corresponding decomposition coefficients (real parameters).
According to our numerical calculation, coefficients $B_i(t), i\in[1,4],$ vanish because of the form of the $H_s$ we choose.
In this case, $H_{s,a}(t)$ reduces to
\begin{equation}
\label{Eq49}
H_{s,a}(t)=A_1(t)\sigma_x\otimes I+A_2(t)I\otimes\sigma_z+A_3(t)\sigma_y\otimes\sigma_z +A_4(t)\sigma_z\otimes\sigma_z.
\end{equation}

\begin{figure}[htbp]
\centering
\includegraphics[width=15cm]{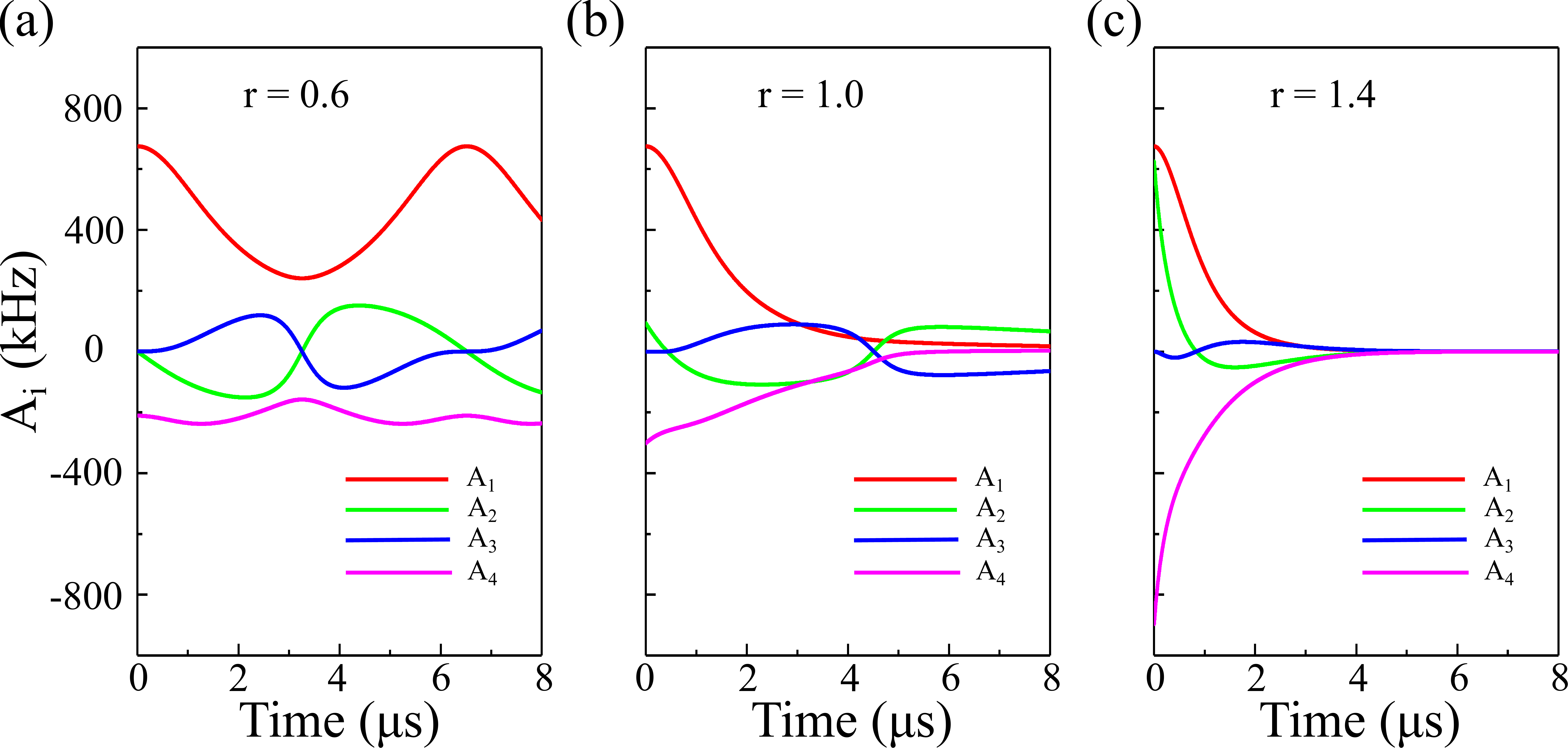}
\caption{\textbf{Parameters $A_i(t)$ in the dilated Hamiltonian.} Parameters $A_i(t)$ ($i=1,2,3,4$) in the dilated Hamiltonian $H_{s,a}(t)$ as a function of t for (a) $r=0.6$, (b) $r=1.0$ and (c) $r=1.4$.}
\label{FigS1}
\end{figure}

Fig.~\ref{FigS1} shows the parameters $A_i(t)$ ($i=1,2,3,4$) of the corresponding dilated Hamiltonian $H_{s,a}(t)$ when $H_s$ is in the region of unbroken $PT$ symmetry (Fig.~\ref{FigS1}.(a)), at the exceptional point (Fig.~\ref{FigS1}.(b)), and in the region of broken $PT$ symmetry (Fig.~\ref{FigS1}.(c)), respectively.

NV center is a kind of point defect in diamond consisting of a substitutional nitrogen atom and an adjoint vacancy.
The electron spin and nuclear spin in NV center consist a highly controllable two-qubit solid-state system.
The electron spin with a spin triple ground state ($S=1$) is coupled with the nearby $^{14}\mathrm{N}$ nuclear spin.
By applying an external magnetic field $B_0$ along the NV axis, the Hamiltonian of NV center can be written as
\begin{equation}
\label{Eq50}
H_{\mathrm{NV}}=2\pi(DS_z^2 + \omega_eS_z + QI_z^2 + \omega_nI_z + AS_zI_z),
\end{equation}
where $\omega_e=-\gamma_eB_0/2\pi$ ($\omega_n=-\gamma_nB_0/2\pi$) is the Zeeman splitting of the  electron ($^{14}\mathrm{N}$ nuclear) spin, with $\gamma_e$ ($\gamma_n$) being the electron ($^{14}\mathrm{N}$ nuclear) gyromagnetic ratio.
$S_z$ and $I_z$ are the spin operators of the electron spin (spin-1) and the $^{14}\mathrm{N}$ nuclear spin (spin-1), respectively.
$D=2.87$ GHz is the axial zero-field splitting parameter for the electron spin.
$Q=-4.95$ MHz is the quadrupole splitting parameter of the $^{14}\mathrm{N}$ nuclear spin.
$A=-2.16$ MHz is the hyperfine coupling parameter.

The experiment is performed in the two qubit subspace spanned by $|m_s=0,m_I=1\rangle, |m_s=0,m_I=0\rangle, |m_s=-1,m_I=1\rangle,$ and $|m_s=-1,m_I=0\rangle$, labeled by $|0\rangle_e|1\rangle_n, |0\rangle_e|0\rangle_n, |-1\rangle_e|1\rangle_n$ and $|-1\rangle_e|0\rangle_n$, in which the Hamiltonian can be simplified as
\begin{equation}
\label{Eq51}
H_0=\pi [-(D-\omega_e-\frac{A}{2})\sigma_z\otimes I+(Q+\omega_n-\frac{A}{2})I\otimes\sigma_z+\frac{A}{2}\sigma_z\otimes\sigma_z].
\end{equation}

To construct $H_{s,a}(t)$ in NV center, we can apply two slightly detuned microwave (MW) pulses to selectively drive the two electron spin transitions, as depicted in Fig.~1b in the main text.
The total Hamiltonian in the two qubit subspace when applying MW pulses can be written as
\begin{equation}
\label{Eq52}
\begin{aligned}
H_{\mathrm{tot}}=\ &\pi [-(D-\omega_e-\frac{A}{2})\sigma_z\otimes I+(Q+\omega_n-\frac{A}{2})I\otimes\sigma_z+\frac{A}{2}\sigma_z\otimes\sigma_z]  \\
&+2\pi\Omega_1(t)\cos[\int_0^t\omega_1(\tau)d\tau+\phi_1(t)]\sigma_x\otimes|1\rangle_n~_n\langle1| +2\pi\Omega_2(t)\cos[\int_0^t\omega_2(\tau)d\tau+\phi_2(t)]\sigma_x\otimes|0\rangle_n~_n\langle0|,
\end{aligned}
\end{equation}
where $\Omega_1(t)$, $\omega_1(t)$ and $\phi_1(t)$ ($\Omega_2(t)$, $\omega_2(t)$ and $\phi_2(t)$) are the Rabi frequency, angular frequency and  phase of the MW pulses which drive the electron spin transition if the nuclear spin is $|1\rangle_n$ ($|0\rangle_n$).
By choosing interaction picture
\begin{equation}
\label{Eq53}
U_{rot}=e^{i\int_0^t[H_0-A_2(\tau)I\otimes\sigma_z-A_4(\tau)\sigma_z\otimes\sigma_z]d\tau},
\end{equation}
the total Hamiltonian transforms to
\begin{equation}
\label{Eq54}
\begin{aligned}
H_{rot}=\ &U_{rot}H_{tot}U_{rot}^\dag-iU_{rot}\frac{dU_{rot}^\dag}{dt}  \\                =\ &A_2(t)I\otimes\sigma_z+A_4(t)\sigma_z\otimes\sigma_z  \\
   &+2\pi\Omega_1(t)\cos[\int_0^t\omega_1(\tau)d\tau+\phi_1(t)]\{\cos[\omega_{MW1}t+\int_0^t2A_4(\tau)d\tau]\sigma_x+\sin[\omega_{MW1}t+\int_0^t2A_4(\tau)d\tau]\sigma_y\}\otimes|1\rangle_n~_n\langle1|  \\
   &+2\pi\Omega_2(t)\cos[\int_0^t\omega_2(\tau)d\tau+\phi_2(t)]\{\cos[\omega_{MW2}t-\int_0^t2A_4(\tau)d\tau]\sigma_x+\sin[\omega_{MW2}t-\int_0^t2A_4(\tau)d\tau]\sigma_y\}\otimes|0\rangle_n~_n\langle0|,
\end{aligned}
\end{equation}
with $\omega_{MW1}$($\omega_{MW2}$) being the energy difference between $|0\rangle_e|1\rangle_n$ and $|-1\rangle_e|1\rangle_n$ ($|0\rangle_e|0\rangle_n$ and $|-1\rangle_e|0\rangle_n$). By choosing
\begin{equation}
\label{Eq55}
\left\{
\begin{aligned}
&\omega_1(t)=\omega_{MW1}+2A_4(t)   \\
&\omega_2(t)=\omega_{MW2}-2A_4(t)   \\
&\Omega_1(t)=\Omega_2(t)=\Omega(t)  \\
&-\phi_1(t)=\phi_2(t)=\phi(t)
\end{aligned},
\right.
\end{equation}
then in the condition of rotating wave approximation, $H_{rot}$ can be reduced to
\begin{equation}
\label{Eq56}
H_{rot}= A_2(t)I\otimes\sigma_z+A_4(t)\sigma_z\otimes\sigma_z
+\pi\Omega(t)\cos[\phi(t)]\sigma_x\otimes I+\pi\Omega(t)\sin[\phi(t)]\sigma_y\otimes I.
\end{equation}
Comparing equation \ref{Eq56} with equation \ref{Eq51}, we can choose
\begin{equation}
\label{Eq57}
\left\{
\begin{aligned}
&\Omega(t)=\frac{\sqrt{A_1^2(t)+A_3^2(t)}}{2\pi},\\
&\phi(t)=\arctan\frac{A_3(t)}{A_1(t)},\\
\end{aligned}
\right.
\end{equation}
to realize Hamiltonian $H_{s,a}(t)$.

\section{\uppercase\expandafter{\romannumeral3}. Characterization of $\psi(t)$}

The state of the system after the evolution governed by the dilated Hamiltonian $H_{s,a}(t)$ is
\begin{equation}
\label{Eq58}
|\Psi(t)\rangle=|\psi(t)\rangle_e|-\rangle_n+\eta_0(t)|\psi(t)\rangle_e|+\rangle_n.
\end{equation}
By applying a rotation of $\pi/2$ along $\mathrm{-X}$ axis on the nuclear spin qubit, as depected in Fig.~1c in the main text, the state evolutes to
\begin{equation}
\label{Eq59}
|\Phi(t)\rangle=|\psi(t)\rangle_e|1\rangle_n+\eta_0(t)|\psi(t)\rangle_e|0\rangle_n.
\end{equation}
Thus the evolution $|\psi(t)\rangle_e$ governed by $\mathcal{PT}$ symmetric Hamiltonian $H_s$ can be characterized by $P_0$, the normalized population of state $|0\rangle_e|1\rangle_n$ after selecting $|1\rangle_n$. $P_0$ can be calculated by $P_0=P_{|0\rangle_e|1\rangle_n}/(P_{|0\rangle_e|1\rangle_n}+P_{|-1\rangle_e|1\rangle_n})$, with $P_{|0\rangle_e|1\rangle_n}$ ($P_{|-1\rangle_e|1\rangle_n}$) being the population of state $|0\rangle_e|1\rangle_n$ ($|-1\rangle_e|1\rangle_n$) of state $|\Phi(t)\rangle$.
This can be acquired by measuring the population distributon of the final state. 

The four levels we choose to perform our experiment give rise to different photoluminescence(PL) rates[\onlinecite{PRB_Steiner}], providing information of the population distribution of the final state. However, different distributions can present the same PL rate, thus a set of pulse sequences are needed to determine the population distribution.

\begin{figure}[htbp]
\centering
\includegraphics[width=15cm]{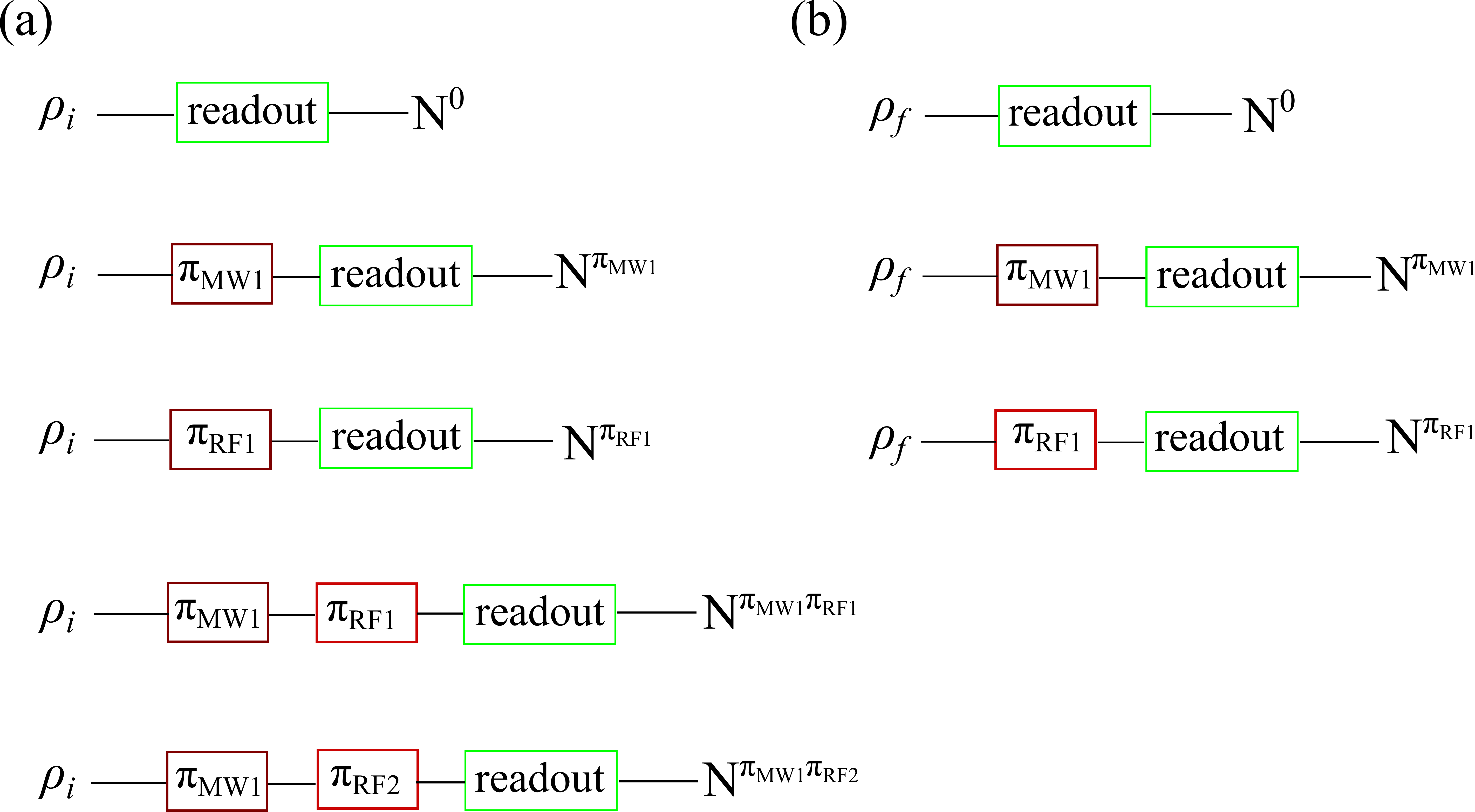}
\caption{
\textbf{Schematic normalization and measurement sequences.}
Here $\rho_\textrm{i}$ denotes the initialized state after a optical excitation, $\rho_\textrm{f}$ denotes the final state after the nuclear $\frac{\pi}{2}$ pulse in readout (main text figure1).
The length of MW pulse ($\pi_{MW1}$) is $896.5~$ns while the length of RF pulse ($\pi_{RF1},\pi_{RF2}$) is $20~\mathrm{\mu s}$.  $N^x$ indicates the detected photoluminescence intensity after applying the pulse sequence $x$.}
\label{FigS2}
\end{figure}

The PL rates of the four levels are measured by the pulse sequences shown in Fig.~\ref{FigS2}(a).
Optical excitation at a static magnetic field of about 506 G can efficiently polarize the system to $|0\rangle_e|1\rangle_n$ via electron spin$-$nuclear spin flip-flop-processes in the electronic excited state of the NV center[\onlinecite{PRL_Jacques}], in combination with an electron-spin dependent relaxation mechanics.
This metnod is used to initialize the system to get the state $\rho_i$ in Fig.~\ref{FigS2}(a).
The number of photons we detect upon PL readout the initialized state is
\begin{equation}
\label{Eq60}
N^0=P_eP_nN_{|0\rangle_e|1\rangle_n}+P_e(1-P_n)N_{|0\rangle_e|0\rangle_n}+(1-P_e)P_nN_{|-1\rangle_e|1\rangle_n}+(1-P_e)(1-P_n)N_{|-1\rangle_e|0\rangle_n},
\end{equation}
where $N_i$ is the number of detected photons if all of the population occupies level $i$, and $P_e$($P_n$) is the population of $|0\rangle_e$ ($|1\rangle_n$) after the optical excitation.
Because of the high polarization of the nuclear spin, $P_n$ is regared as 1 during the calculation.
Considering all the pulse sequences in Fig.~\ref{FigS2}(a), we can obtain the following equations
\begin{equation}
\label{Eq61}
\begin{bmatrix}
P_e & 0 & 1-P_e & 0 \\
1-P_e & 0 & P_e & 0 \\
0 & P_e & 1-P_e & 0 \\
0 & 1-P_e & P_e & 0 \\
1-P_e & 0 & 0 & P_e \\
\end{bmatrix}
\begin{bmatrix}
 N_{|0\rangle_e|1\rangle_n} \\ N_{|0\rangle_e|0\rangle_n} \\ N_{|-1\rangle_e|1\rangle_n} \\ N_{|-1\rangle_e|0\rangle_n} \\
 \end{bmatrix}
= \begin{bmatrix}
N^0 \\ N^{\pi_{MW1}} \\ N^{\pi_{RF1}} \\ N^{\pi_{MW1}\pi_{RF1}} \\ N^{\pi_{MW1}\pi_{RF2}} \\ \end{bmatrix},
\end{equation}
where $N^x$ indicates the detected PL after applying the pulse sequence $x$. $\pi_{MW1}$ ($\pi_{RFR}$,$\pi_{RF2}$) denotes a selective $\pi$ pulse between $|0\rangle_e|1\rangle_n$ and $|-1\rangle_e|1\rangle_n$ ($|0\rangle_e|1\rangle_n$ and $|0\rangle_e|0\rangle_n$, $|-1\rangle_e|1\rangle_n$ and $|-1\rangle_e|0\rangle_n$), which flips the poulations in these levels.
By solving equation \ref{Eq61}, the PL rate of each level can be obtained.

Knowing $N_i$ we can determine the occupation probabilits $P_i$ of the final state $\rho_f$ by pulse sequences shown in Fig.~\ref{FigS2}(b).
The PL of an final state with level occupation probabilities $P_i$ is
\begin{equation}
\label{Eq62}
N^0=P_{|0\rangle_e|1\rangle_n}N_{|0\rangle_e|1\rangle_n}+P_{|0\rangle_e|0\rangle_n}N_{|0\rangle_e|0\rangle_n}+P_{|-1\rangle_e|1\rangle_n}N_{|-1\rangle_e|1\rangle_n}+P_{|-1\rangle_e|0\rangle_n}N_{|-1\rangle_e|0\rangle_n}.
\end{equation}
By flipping populations within the two-qubit subspace using the pulse sequences in Fig.~\ref{FigS2}(b) and measuring the resulting PL we can calculate the $P_i$ from
\begin{equation}
\label{Eq63}
\begin{bmatrix}
N_{|0\rangle_e|1\rangle_n} & N_{|0\rangle_e|0\rangle_n} & N_{|-1\rangle_e|1\rangle_n} & N_{|-1\rangle_e|0\rangle_n} \\
N_{|-1\rangle_e|1\rangle_n} & N_{|0\rangle_e|0\rangle_n} & N_{|0\rangle_e|1\rangle_n} & N_{|-1\rangle_e|0\rangle_n} \\
N_{|0\rangle_e|0\rangle_n} & N_{|0\rangle_e|1\rangle_n} & N_{|-1\rangle_e|1\rangle_n} & N_{|-1\rangle_e|0\rangle_n} \\
1      &  1     &      1 & 1      \\
\end{bmatrix}\begin{bmatrix} P_{|0\rangle_e|1\rangle_n} \\ P_{|0\rangle_e|0\rangle_n} \\ P_{|-1\rangle_e|1\rangle_n} \\ P_{|-1\rangle_e|0\rangle_n} \\ \end{bmatrix} = \begin{bmatrix} N^0 \\ N^{\pi_{MW1}} \\ N^{\pi_{RF1}} \\ 1 \\ \end{bmatrix}.
\end{equation}

\section{\uppercase\expandafter{\romannumeral4}. Experimental acquisition of the eigenvalues of $\mathcal{PT}$ symmetric Hamiltonians}

By fitting the experimental evolution curve to the theoretical evolution curve we can get the parameter $r_{exp}$, then the eigenvalues of $\mathcal{PT}$ symmetric Hamiltonian $H_s$ can be calculated by $E_\pm=\pm\sqrt{1-r_{exp}^2}$.
The Hamiltonian we realized in our experimental system is
\begin{equation}
\label{Eq64}
H_s=
\begin{pmatrix}
& ir  &1             \\
&1             &-ir  \\
\end{pmatrix}.
\end{equation}
The time evolution operator corresponding to $H_s$ takes the form
\begin{equation}
\label{Eq65}
\begin{aligned}
U(t)&=e^{-iH_st}  \\
&=
\begin{pmatrix}
&\frac{e^{t\sqrt{r^2-1}}(r+\sqrt{r^2-1}) -e^{-t\sqrt{r^2-1}}(r-\sqrt{r^2-1})}{2\sqrt{r^2-1}}
&\frac{ie^{-t\sqrt{r^2-1}}-ie^{t\sqrt{r^2-1}}}{2\sqrt{r^2-1}}   \\
&\frac{ie^{-t\sqrt{r^2-1}}-ie^{t\sqrt{r^2-1}}}{2\sqrt{r^2-1}}
&\frac{e^{t\sqrt{r^2-1}}(-r+\sqrt{r^2-1}) -e^{-t\sqrt{r^2-1}}(-r-\sqrt{r^2-1})}{2\sqrt{r^2-1}}                                                                \\
\end{pmatrix}.
\end{aligned}
\end{equation}
The initial state we choose is $|\psi(0)\rangle=|0\rangle=(1,0)^T$. Then the final state at time $t$, after the evolution governed by Hamiltonian $H_s$, is
\begin{equation}
\label{Eq66}
|\psi(t)\rangle=\frac{1}{2\sqrt{r^2-1}}
\begin{pmatrix}
&e^{t\sqrt{r^2-1}}(r+\sqrt{r^2-1})-e^{-t\sqrt{r^2-1}}(r-\sqrt{r^2-1})    \\
&ie^{-t\sqrt{r^2-1}}-ie^{t\sqrt{r^2-1}}   \\	
\end{pmatrix}.
\end{equation}
So the population of state $|0\rangle$ at moment $t$ is
\begin{equation}
\label{Eq67}
P_0=\frac{|e^{t\sqrt{r^2-1}}(r+\sqrt{r^2-1}) -e^{-t\sqrt{r^2-1}}(r-\sqrt{r^2-1})|^2}{|e^{t\sqrt{r^2-1}}(r+\sqrt{r^2-1}) -e^{-t\sqrt{r^2-1}}(r-\sqrt{r^2-1})|^2 +|ie^{-t\sqrt{r^2-1}}-ie^{t\sqrt{r^2-1}}|^2}.
\end{equation}
We use formula \ref{Eq67} to fit our experimental data to get the parameter $r_{exp}$.
The fitting result is shown on TABLE \ref{table1}.

\begin{table}[!h]\centering
\caption{\textbf{Parameter $r_{\mathrm{exp}}$ obtained from the time evolution under $\mathcal{PT}$ symmetric Hamiltonian.} $r$ is the parameter at which the experiment implemented. $r_{\mathrm{exp}}$ is obtained by fitting the evolution curve. $\delta r_{\mathrm{exp}}$ is the fitting error.}
\textrm{\\}
\begin{tabular}{|c|c|c|c|c|c|c|c|c|c|c|c|c|c|c|c|c|}
\hline
$r$ & 0 & 0.1 & 0.2 & 0.3 & 0.4 & 0.5 & 0.6 & 0.7 & 0.8 & 0.9 & 1.0 & 1.1 & 1.2 & 1.3 & 1.4 & 1.5 \\
\hline
$r_{\mathrm{exp}}$  &  0.006  &  0.099 & 0.191 & 0.328 & 0.416 & 0.472 & 0.616 & 0.713 & 0.800 & 0.906 & 1.002 & 1.079 & 1.170 & 1.321 & 1.418 & 1.509 \\
\hline
$\delta r_{\mathrm{exp}}$ & 0.018 & 0.024 & 0.014 & 0.016 & 0.009 & 0.015 & 0.006 & 0.006 & 0.003 & 0.006 & 0.010 & 0.015 & 0.021 & 0.019 & 0.038 & 0.001 \\
\hline
\end{tabular}
\label{table1}
\end{table}

\section{\uppercase\expandafter{\romannumeral5}. $T_2^\star$ of NV center}

\begin{figure}[htbp]
\centering
\includegraphics[width=15cm]{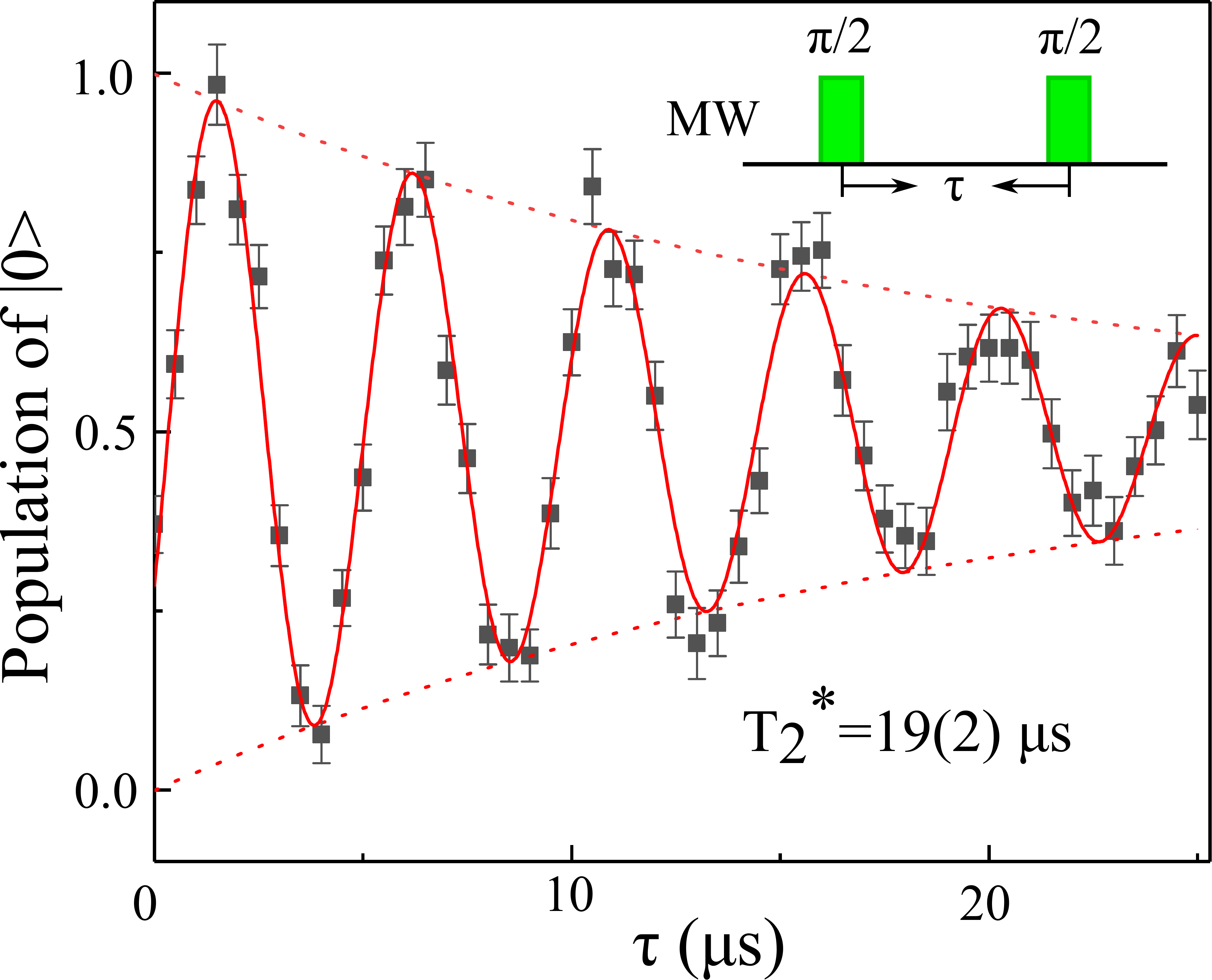}
\caption{\textbf{Coherence time of NV center.}   Result of the Ramsey experiment (insert, pulse sequence) for the electron spin. The solid red line is fit to the experiment data (black square), the red dashed line is the fit to the envelope curve. The decay time of FID is measured to be $T_2^\star=19(2)~\mathrm{\mu s}$.}
\label{FigS3}
\end{figure}

The sample used in our experiment is isotopically purified ([$^{12}$C]=99.9\%). So the coherence time of the electron spin, which is the system qubit in this experiment, is prolonged.
Ramsey sequence is utilized to measure the $T_2^\star$ of the electron spin, and the result is demonstrated in Fig.~\ref{FigS3}.
The measurement shows that the $T_2^\star$ of our sample is $19(2)~\mathrm{\mu s}$. 

\section{\uppercase\expandafter{\romannumeral6}. Nuclear spin qubit operation}

\begin{figure}[htbp]
\centering
\includegraphics[width=15cm]{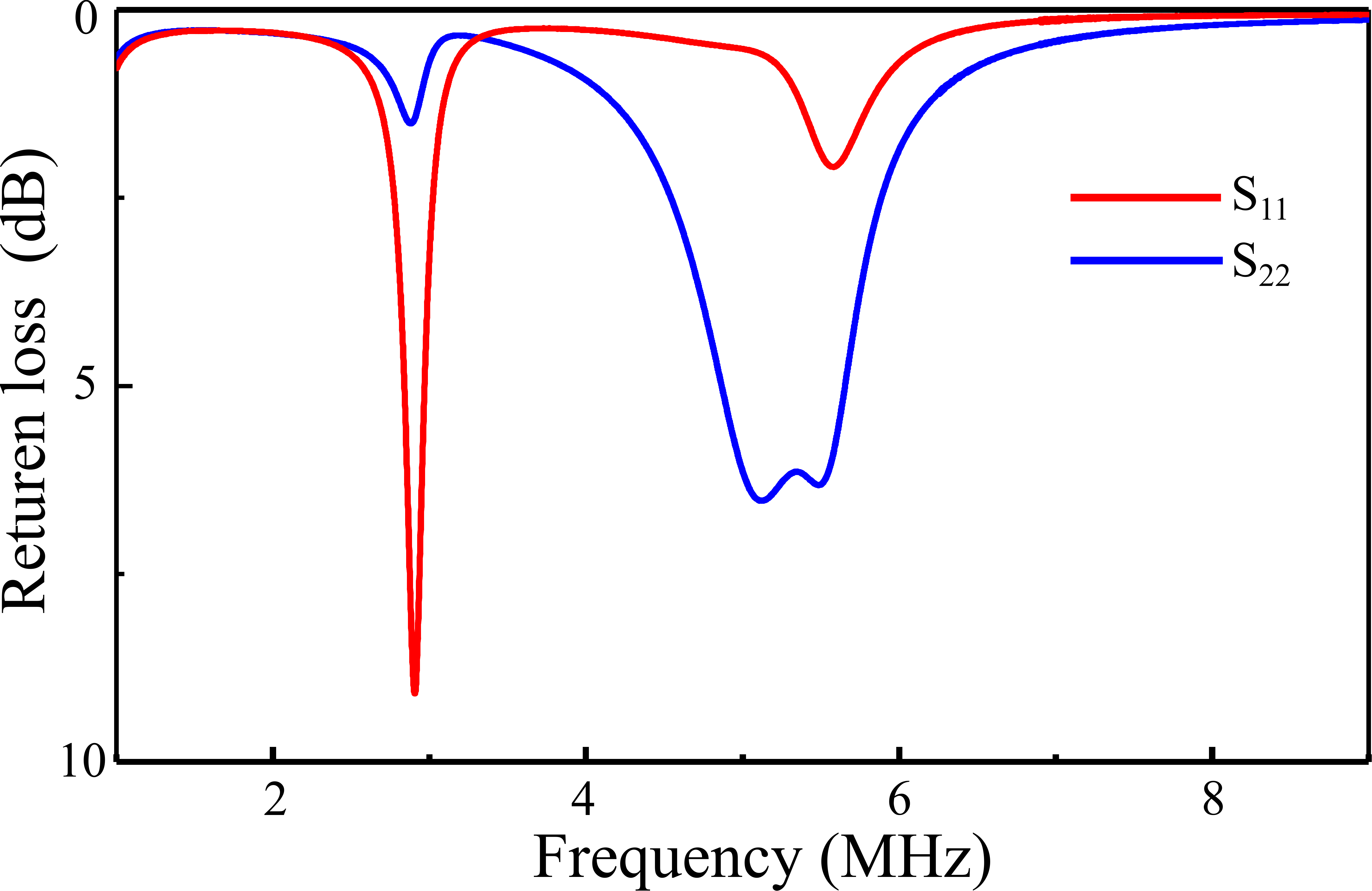}
\caption{\textbf{Return loss of the RF Coil.}
Return loss of the RF coil, red line stands for the low frequency port, at $2.9~\mathrm{MHz}$, the return loss is $-9.1~\mathrm{dB}$, bule line stands for the high frequency port, at $5.1~\mathrm{MHz}$, the retirn loss is $-6.5~\mathrm{dB}$. }
\label{FigS4}
\end{figure}

The single qubit operations of the nuclear spin qubit in the state prepartion and read out are realized by applying two channel radio frequency (RF) pulses simultaneously.
The frequencies of the RF pulses are $2.9~\mathrm{MHz}$ and $5.1~\mathrm{MHz}$, corresponding to the nuclear spin transition frequencies.
The RF pulses are generated by AWG and then carried by a homebuild RF coil with dual resonance frequencies.
The coil is designed according to ref.[\onlinecite{Dissertation_Dabirzadeh}]. There are two input ports at the coil to input the RF pulses, and the return loss of the two ports is shown in Fig.~\ref{FigS4}.
The Rabi frequencies of the nuclear spin transitions at different electron subspace are both calibrated to be 25~$\mathrm{kHz}$.

\end{document}